\documentclass{aa}  
\usepackage{graphicx}
\usepackage[varg]{txfonts}
\begin{document}

   \title{Do Miras show the long secondary periods?}
   \titlerunning{Do Miras show the LSP?}

   \author{Micha{\l} Pawlak}
          
    \authorrunning{M. Pawlak}

   \institute{Astronomical Observatory, Jagiellonian University, ul. Orla 171, 30-244 Krak{\'o}w, Poland\\
              \email{michal1.pawlak@uj.edu.pl}
             }

   \date{}

 
  \abstract
   {}
   {The long secondary period (LSP) phenomenon, which is commonly observed in pulsating red giants, has not been detected in any Mira yet. The goal of this paper is to verify, if there is a physical reason for this or if it is simply an observational bias.}
   {The OGLE-III Sample of Long Period Variables in the Large Magellanic Cloud, containing 1663 Miras, is used to perform a search for secondary periodicity in these objects and identify candidates for the long secondary period stars based on the location on the period-luminosity diagram.}
   {Out of 1663 Miras, 108 were identified as potential candidates, with variability broadly consistent with LSP. This makes 7\% of the whole Mira sample in the Large Magellanic Cloud. Most, if not all of the Mira LSP candidates are C-rich stars.}
   {The results of this analysis  suggest that Miras may exhibit long secondary periods. However, the long-term variability can also be related to period and amplitude irregularities that Miras are known to exhibit. Further study will be necessary to draw a definitive conclusion.}


   \maketitle
%

\section{Introduction}

Long period variables (LPVs) are red giant branch (RGB) and asymptotic giant branch (AGB) stars that show regular or semi-regular, periodic variability of the luminosity due to radial, often multi-modal pulsations. Apart from pulsations, a large number of LPVs show an additional type of variability known as the long secondary periods (LSP). The term LSP has been introduced by Wood et al. (1999), however, the phenomenon itself has been known before \citep{oconnell1933, paynegaposchkin1954, houk1963}. \citet{soszynskietal2007} estimate that between 25\% and 50\% of LPVs in OGLE Collection of the Variable Stars (OCVS) in  the Magellanic Clouds show LSP. The LSP was also observed in 26\% of the LPVs in the ASAS-SN/APOGEE catalog of variable stars (Pawlak et al. 2019).

Despite the large number of known LSP stars and the fact that the phenomenon has been known for decades, the mechanism behind it still remains unclear. The two most likely hypotheses on the origin of LSP are either non-radial pulsation modes \citep{wood2000a,wood2000b,hinkle2002,wood2004,saio2015} or binarity with a low-mass, sub-stellar companion - either a brown dwarf or a large exoplanet \citep{wood1999,soszynski2007,soszynski2014,soszynski2021}.

The OCVS distinguishes three classes of LPVs: OSARGs, SRVs, and Miras. The vast majority of the known LSP stars belong to the first class, while some of the SRVs are also known to show LSP. However, none of the known Mira variables have been confirmed to have LSP.

The lack of LSP Miras may have a physical reason. Miras are very specific objects in terms of their evolutionary status and variability pattern, therefore some of their characteristics may impede the onset of LSP variability. This possibility is even harder to exclude, due to the fact, that the mechanism behind LSP is still not fully explained. For instance, if LSP is tied specifically to the first-overtone pulsation \citep{trabucchi2017, mcdonald2019}, or a transition between different pulsation modes \citep{pawlak2021}, it will naturally not appear in Miras, which are fundamental-mode pulsators.

However, it is also possible that the lack of LSP detection in Miras is simply an observational bias. As Miras typically have very high amplitudes of the fundamental mode pulsations, it is likely that any other type of variability present in them would be obscured by the principal source of variability.

It should also be noted, that while no clear Mira LSP has been identified yet, the existence of long-timescale variability in Miras is well known. The irregularities of periods in Miras were already studied by \citet{eddington1929}. Significant changes in periods of individual Miras were detected by \citet{plakidis1932} and \citet{sterne1937}. More recent studies show that Miras can undergo changes in period, mean magnitude, and amplitude \citep{percy1999a,percy1999b}.

The goal of this paper is to investigate, whether the lack of LSP in Miras is caused by a physical reason or an observational bias. The structure of the paper is as follows. In Sec.~2 I describe the data set that I use, and the analysis I perform. I discuss the results in Sec.~3 and summarize them in Sec.~4.

\section{Analysis}

For the purpose of this study, I use the OGLE-III catalog of LPVs in the Large Magellanic Cloud \citep{soszynskietal2007}. The technical details of the OGLE-III survey can be found in \citet{udalski2003}. The analyzed sample contains 1663 Miras, none of them marked as an LSP star. At the same time, there are $6\;586$ and 365 stars flagged as LSP among the $78\;919$ OSARGs and $11\;123$ SRVs, respectively. The selection of Miras in the OGLE catalog was based on the method described in \citet{soszynski2005}. The criterion that was used was the location of the star on the near-infrared PL diagram and the amplitude higher than 2.5~mag or 0.9~mag in $V$ and $I$-band, respectively. The final classification was verified by visual inspection of the light curves.

The fact that LSP is very often detected in low-amplitude OSARGs, but only in a small number of higher-amplitude SRVs suggests, that a selection bias might be present. Pulsational variability in SRVs has amplitudes that are often comparable to or higher than the typical amplitudes of the LSP. Therefore, LSP is likely to remain undetected in many cases, as it is obscured by the pulsational variability. This bias can potentially be even stronger in Miras, where the fundamental-mode pulsations typically have amplitudes of a few magnitudes. These would practically always be significantly higher than the amplitude of putative LSP variability, which rarely exceeds 1~mag.

\begin{figure*}[ht]
\centering
	\includegraphics[width=0.99\textwidth]{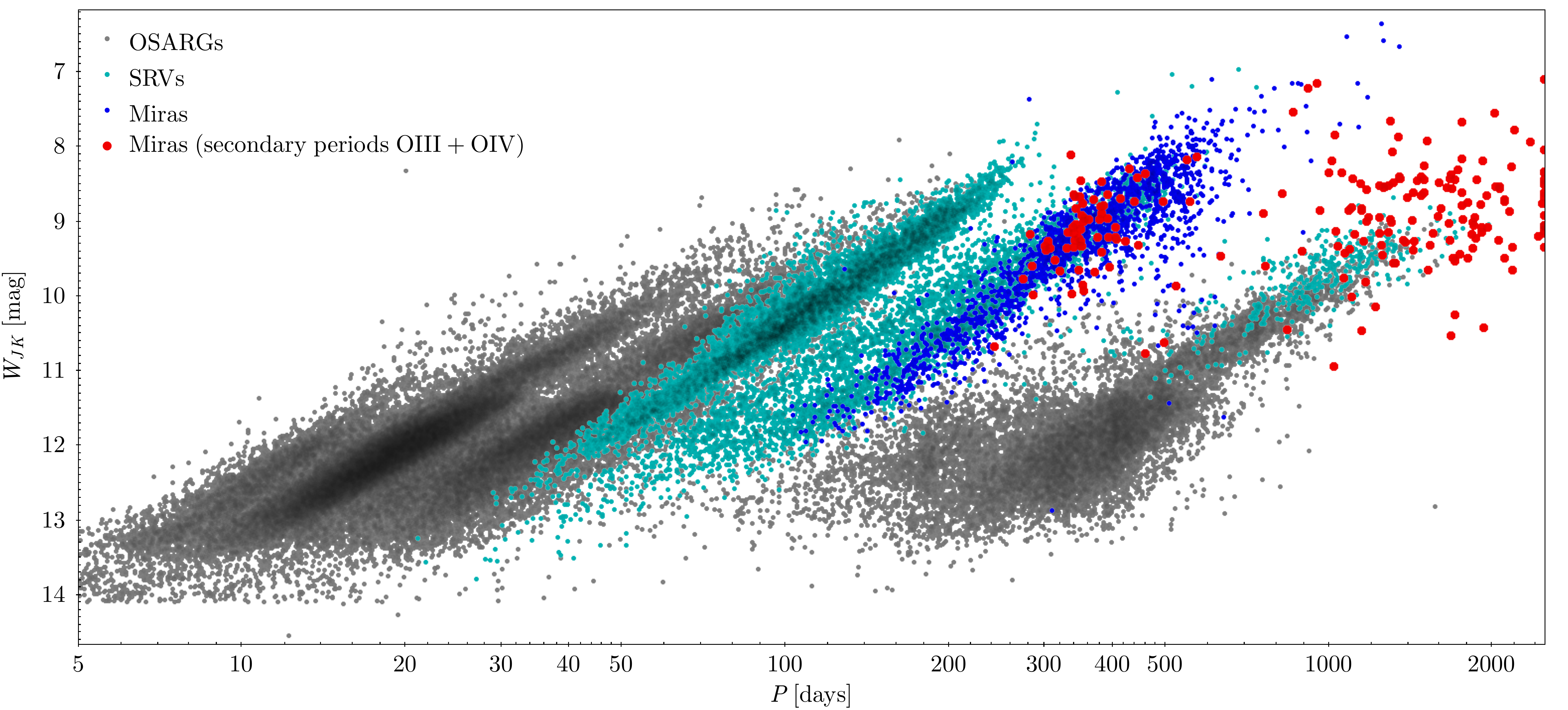} 
	\caption{PLR in the reddening-free  in $W_{JK}$ index, formed by the three types of LPVs: OSARGs, SRVs, and Miras as classified by \citet{soszynskietal2007}. Miras, for which a period with a significantly high S/N in the prewhitened light curve was detected, are marked in red, with the strongest period detected after prewhitening.}  
	    \label{fig:pl_wjk_new.eps}
\end{figure*}

\begin{figure*}
\label{lcs}
 \includegraphics[width=0.26\textwidth]{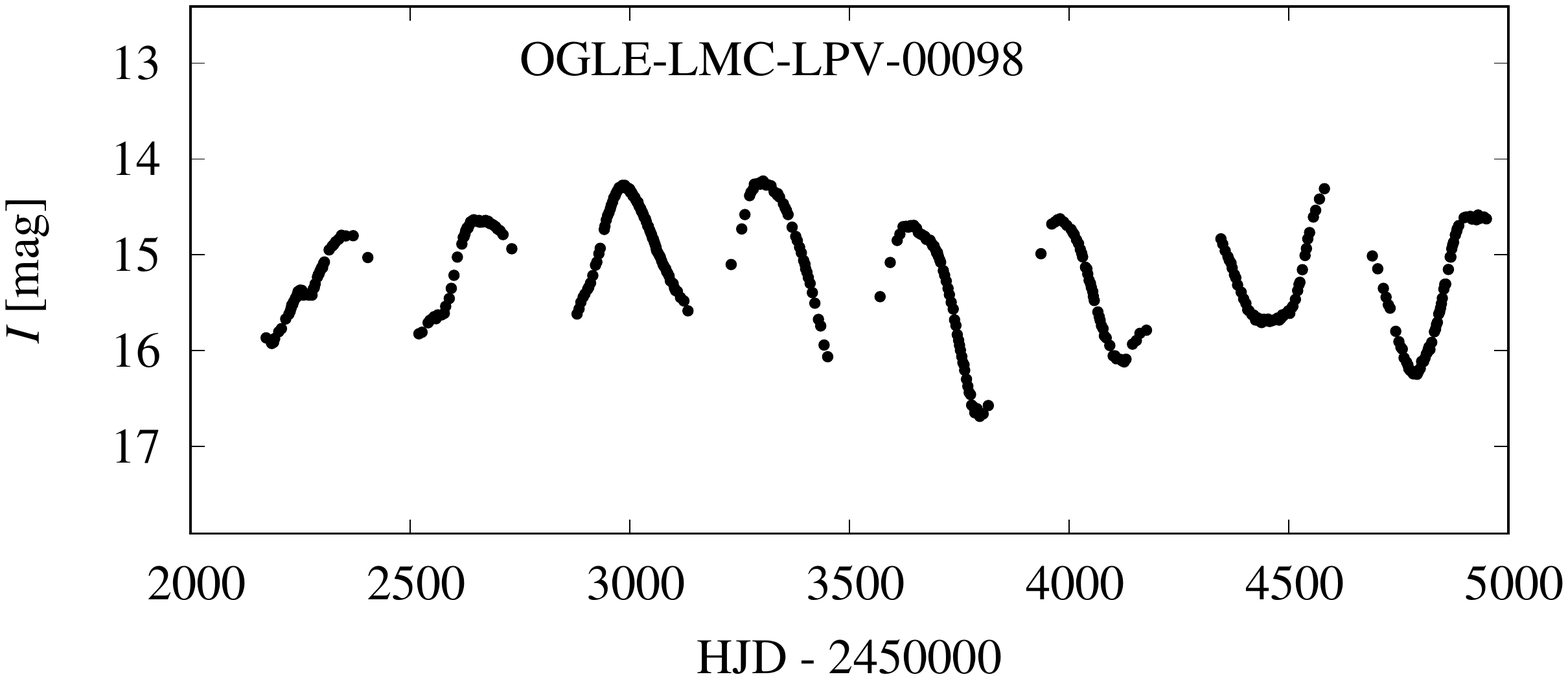} \includegraphics[width=0.26\textwidth]{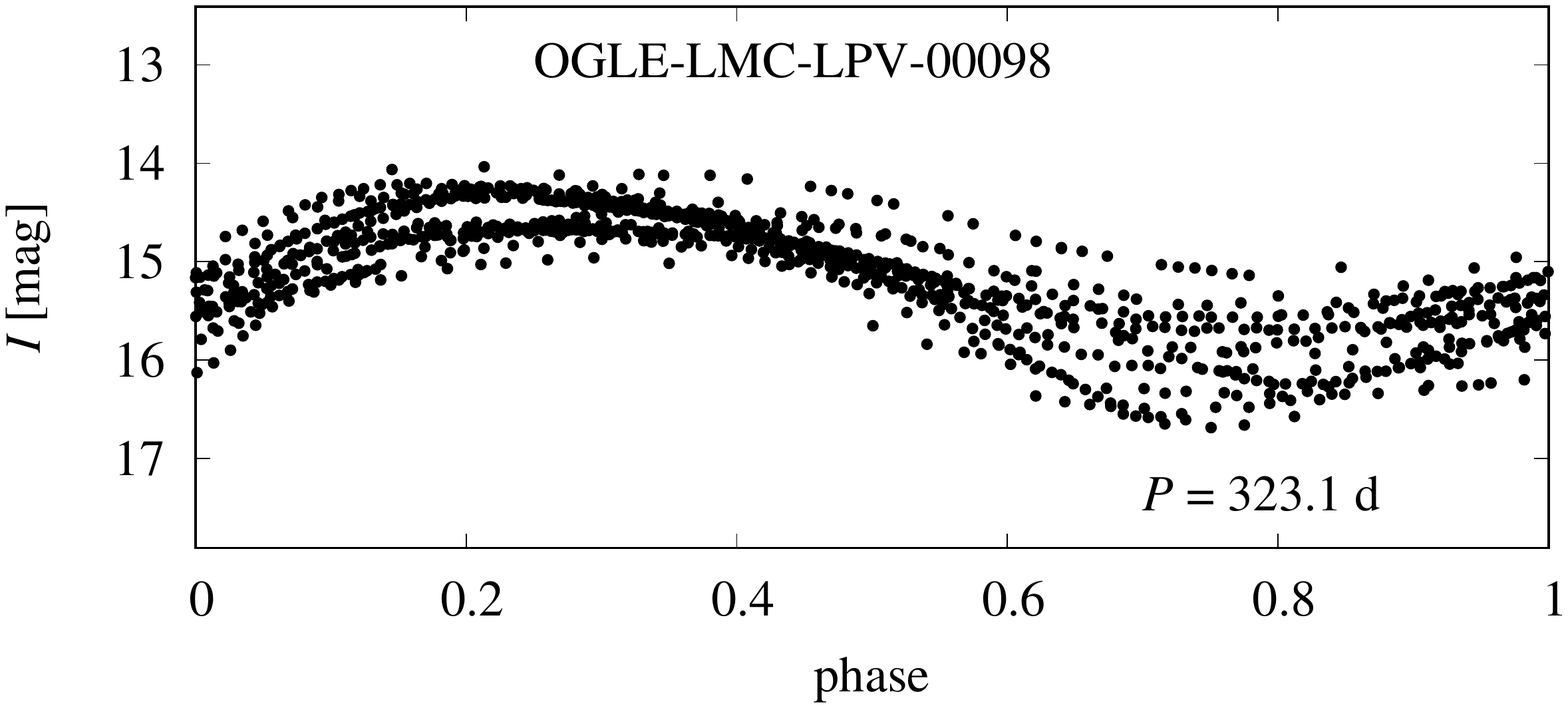}  \includegraphics[width=0.26\textwidth]{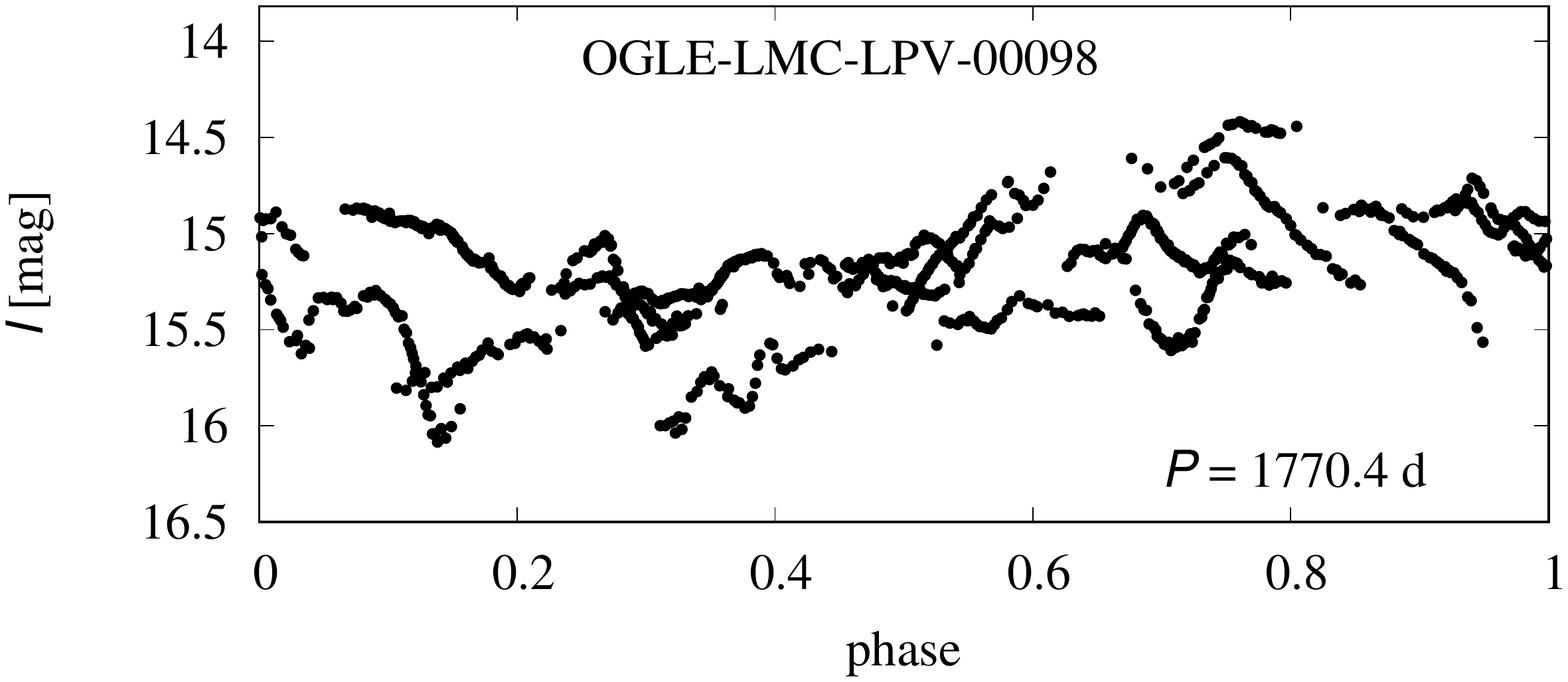}\\
 \includegraphics[width=0.26\textwidth]{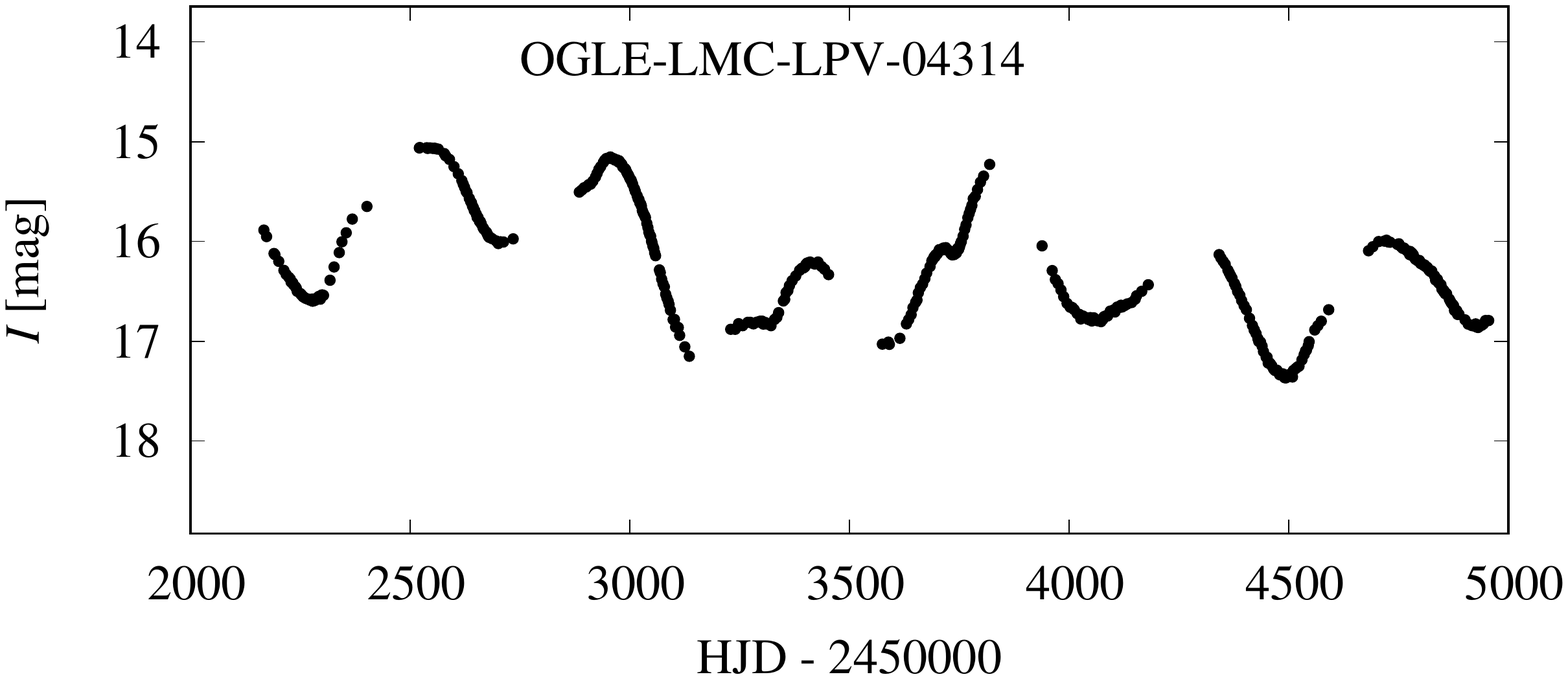} \includegraphics[width=0.26\textwidth]{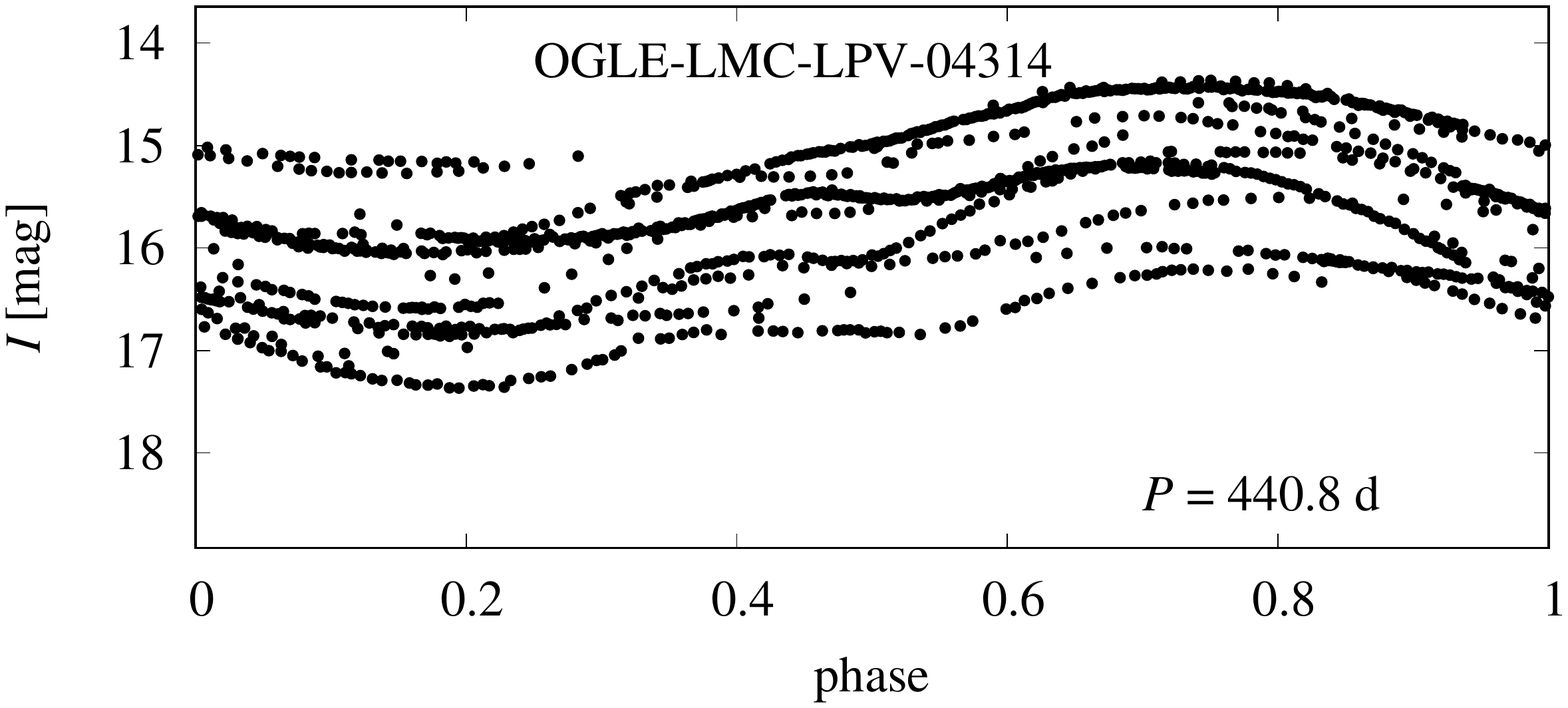}  \includegraphics[width=0.26\textwidth]{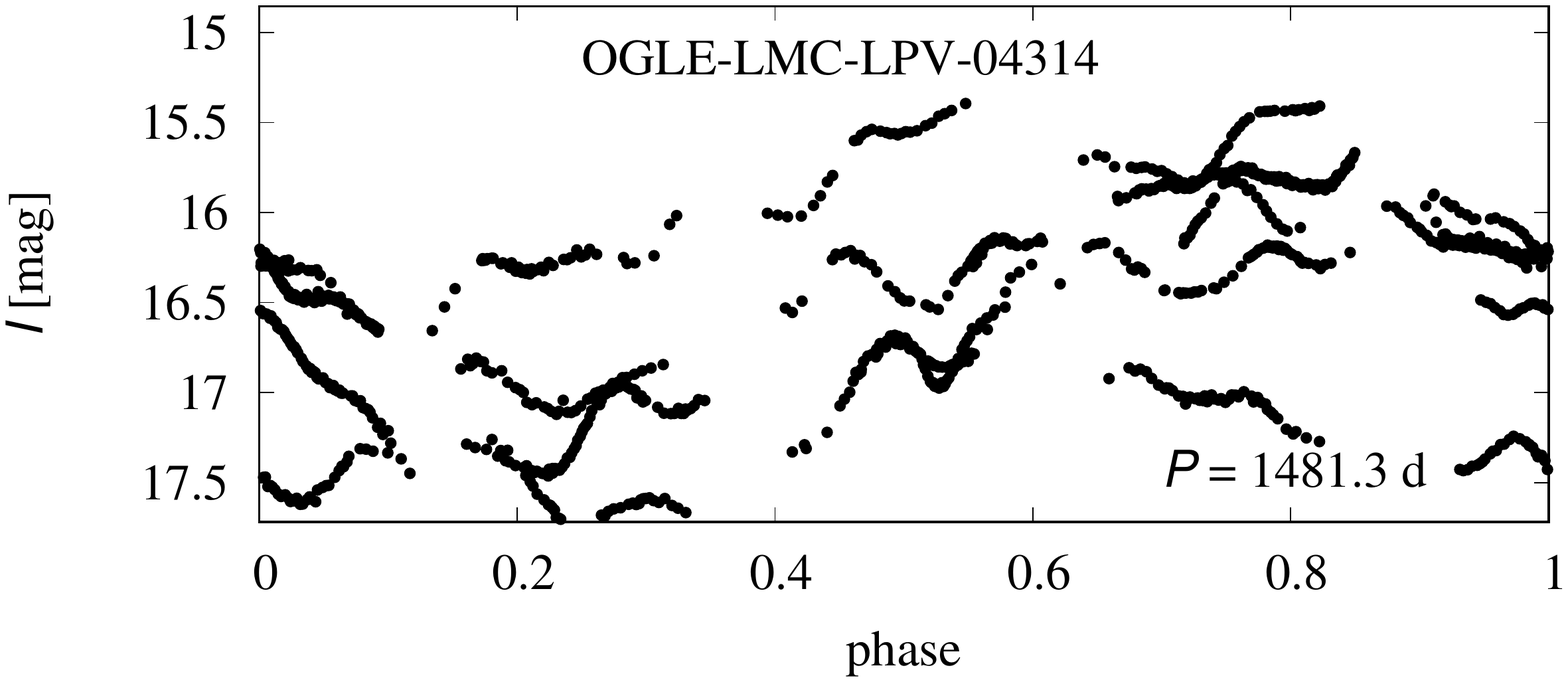}\\
 \includegraphics[width=0.26\textwidth]{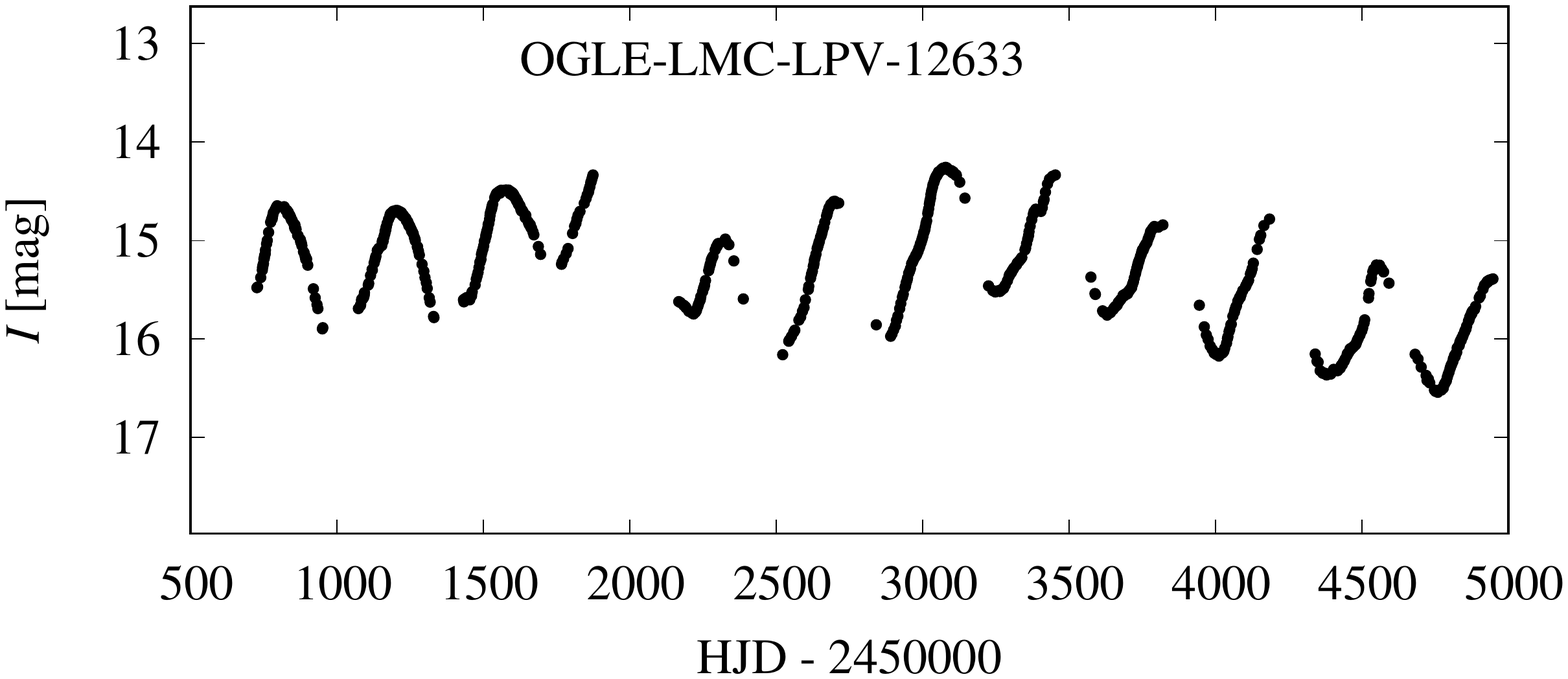} \includegraphics[width=0.26\textwidth]{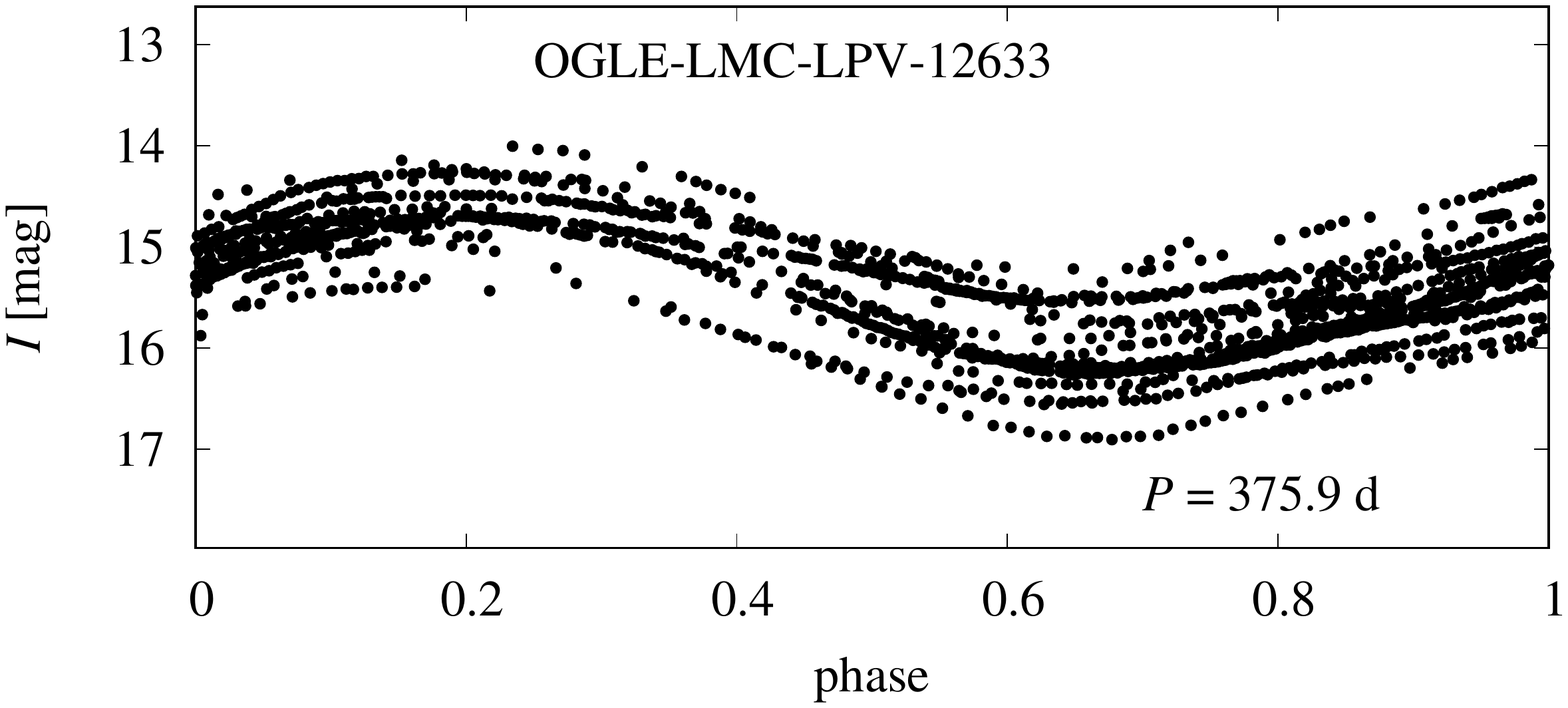}  \includegraphics[width=0.26\textwidth]{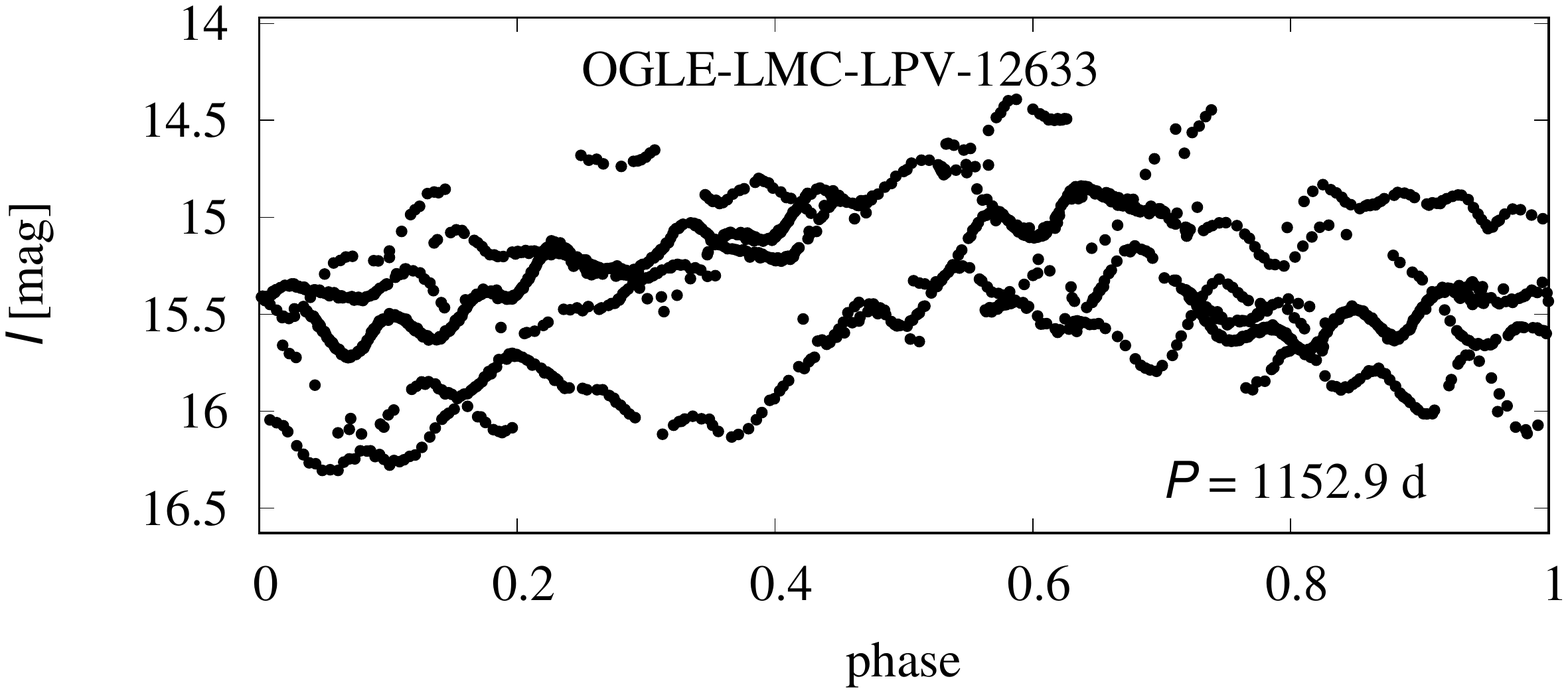}\\
 \includegraphics[width=0.26\textwidth]{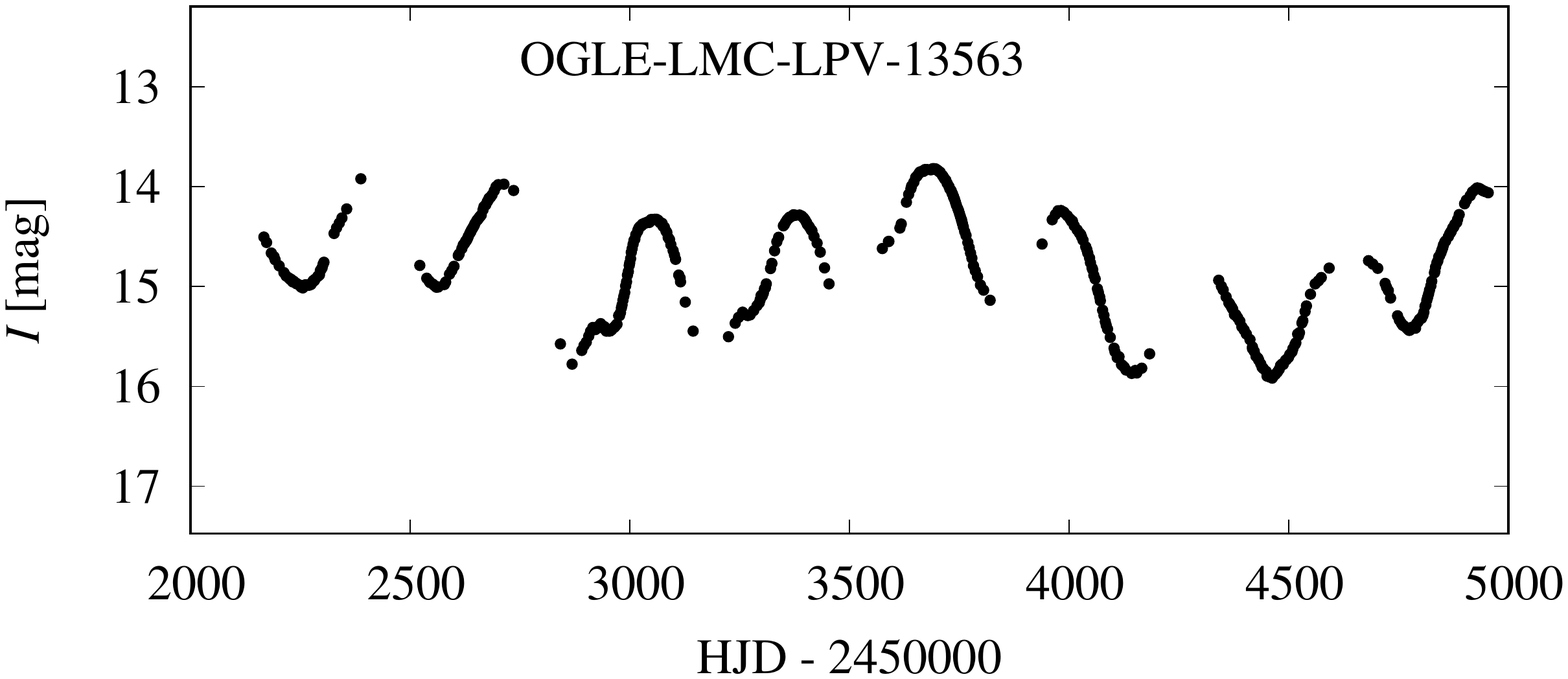}
 \includegraphics[width=0.26\textwidth]{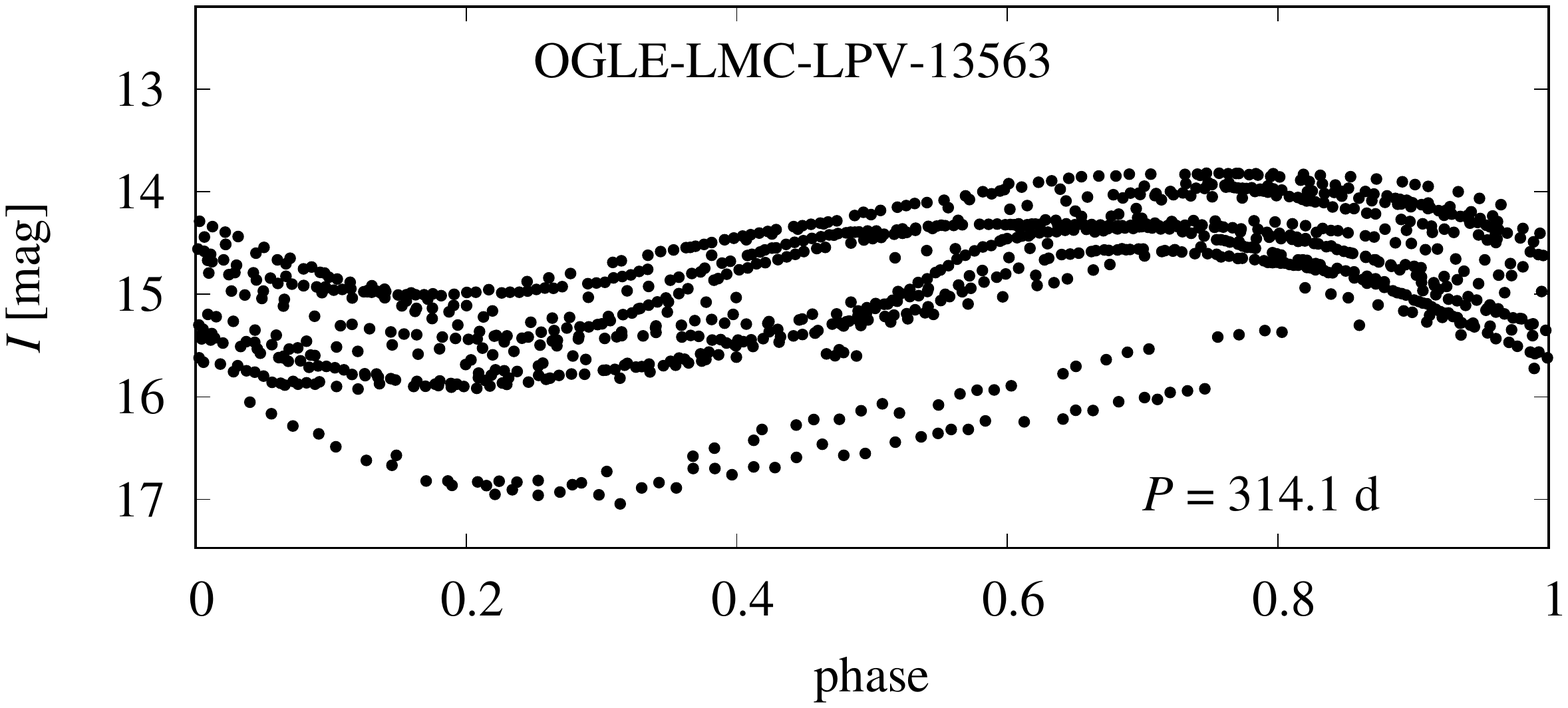}
 \includegraphics[width=0.26\textwidth]{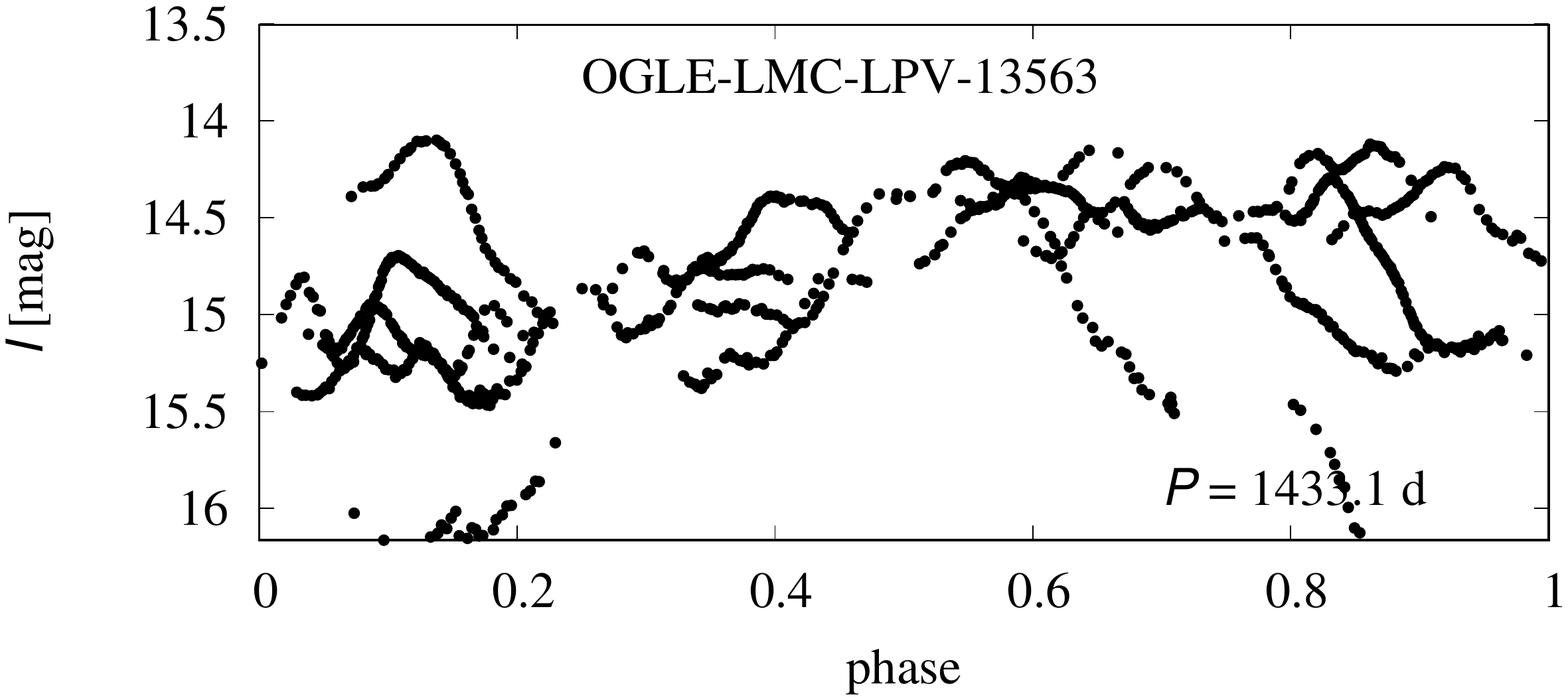}\\
 \includegraphics[width=0.26\textwidth]{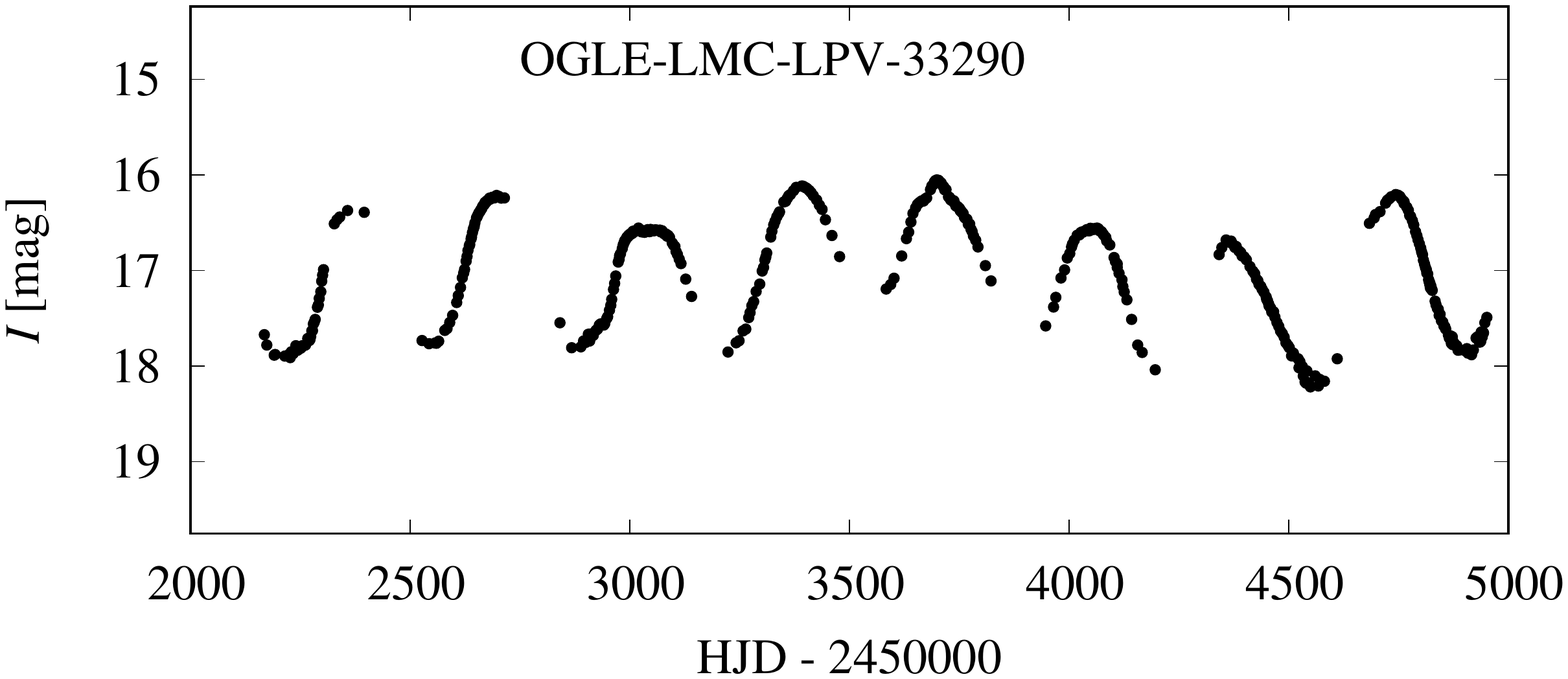} \includegraphics[width=0.26\textwidth]{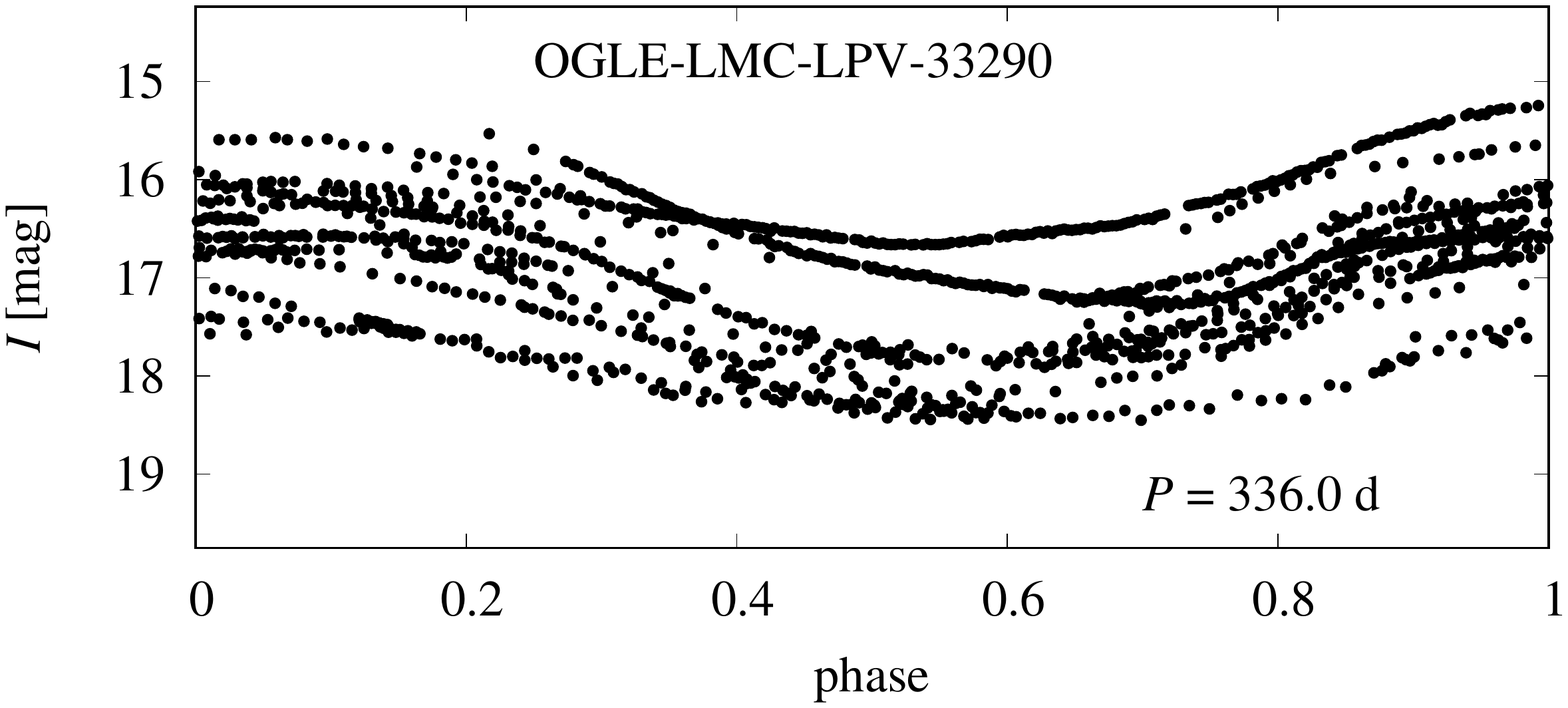}  \includegraphics[width=0.26\textwidth]{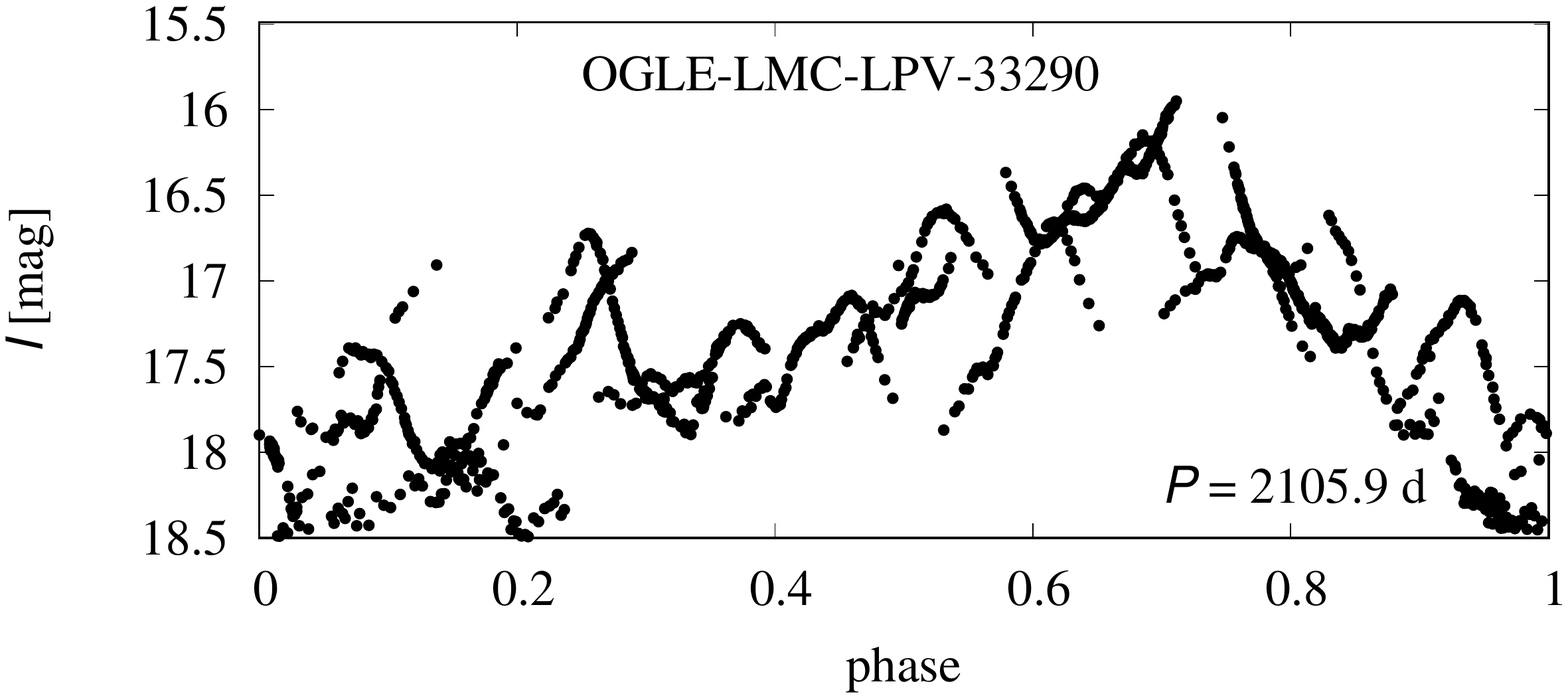}\\
 \includegraphics[width=0.26\textwidth]{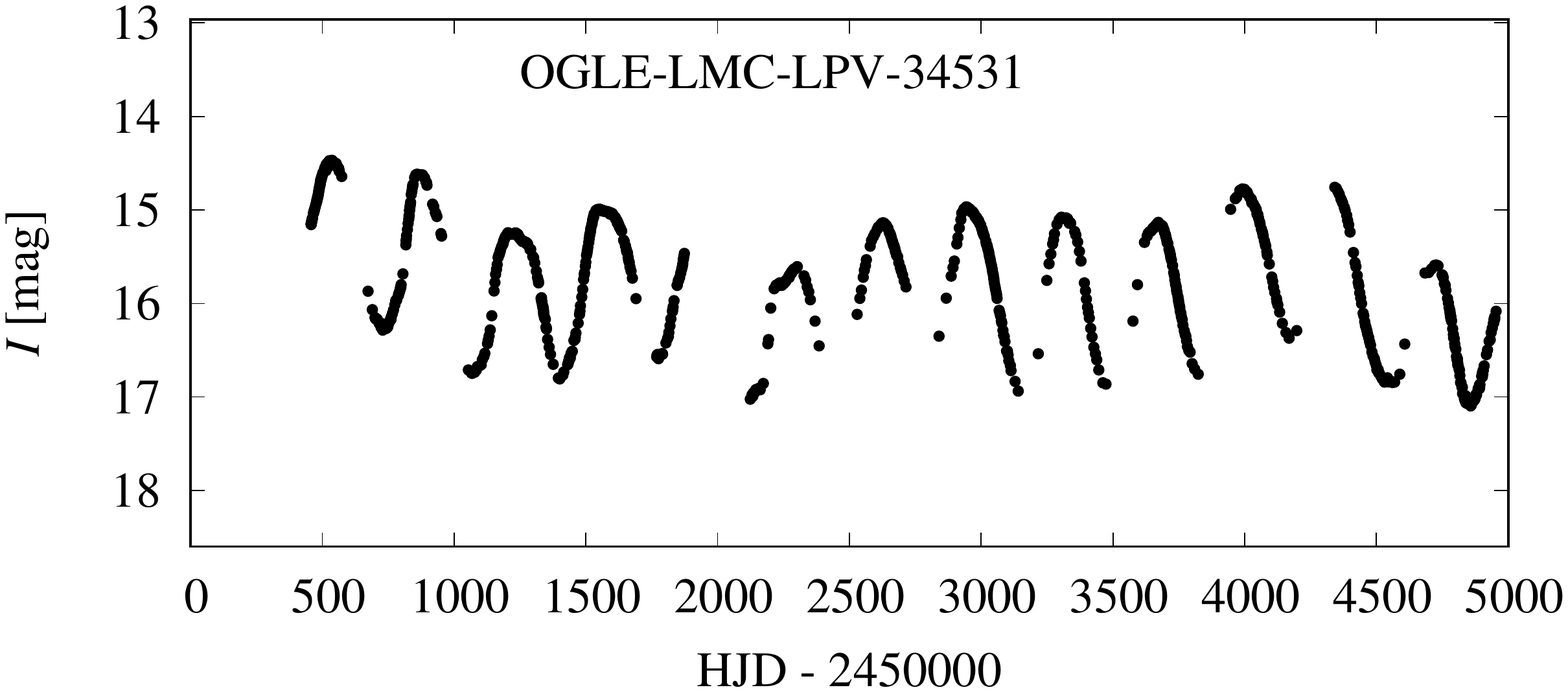} \includegraphics[width=0.26\textwidth]{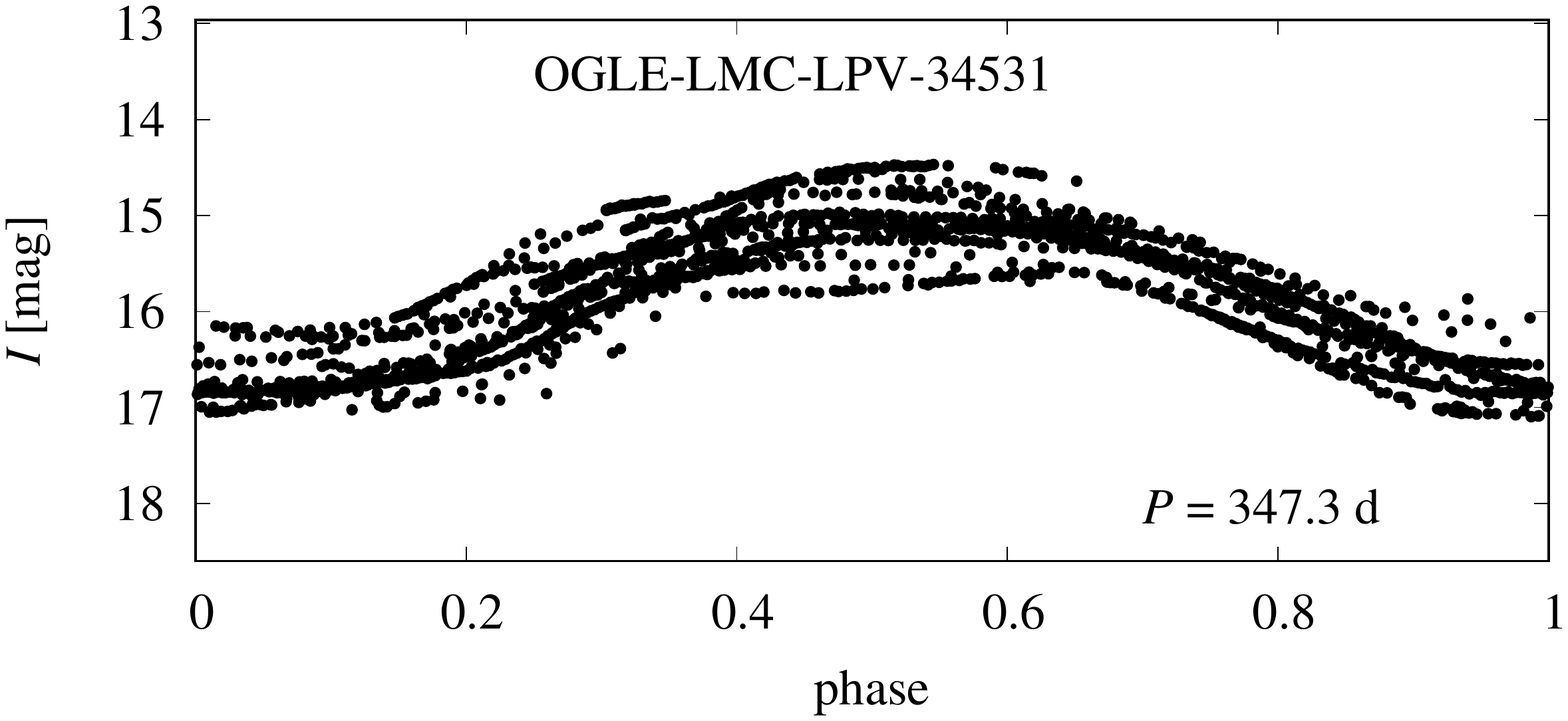}  \includegraphics[width=0.26\textwidth]{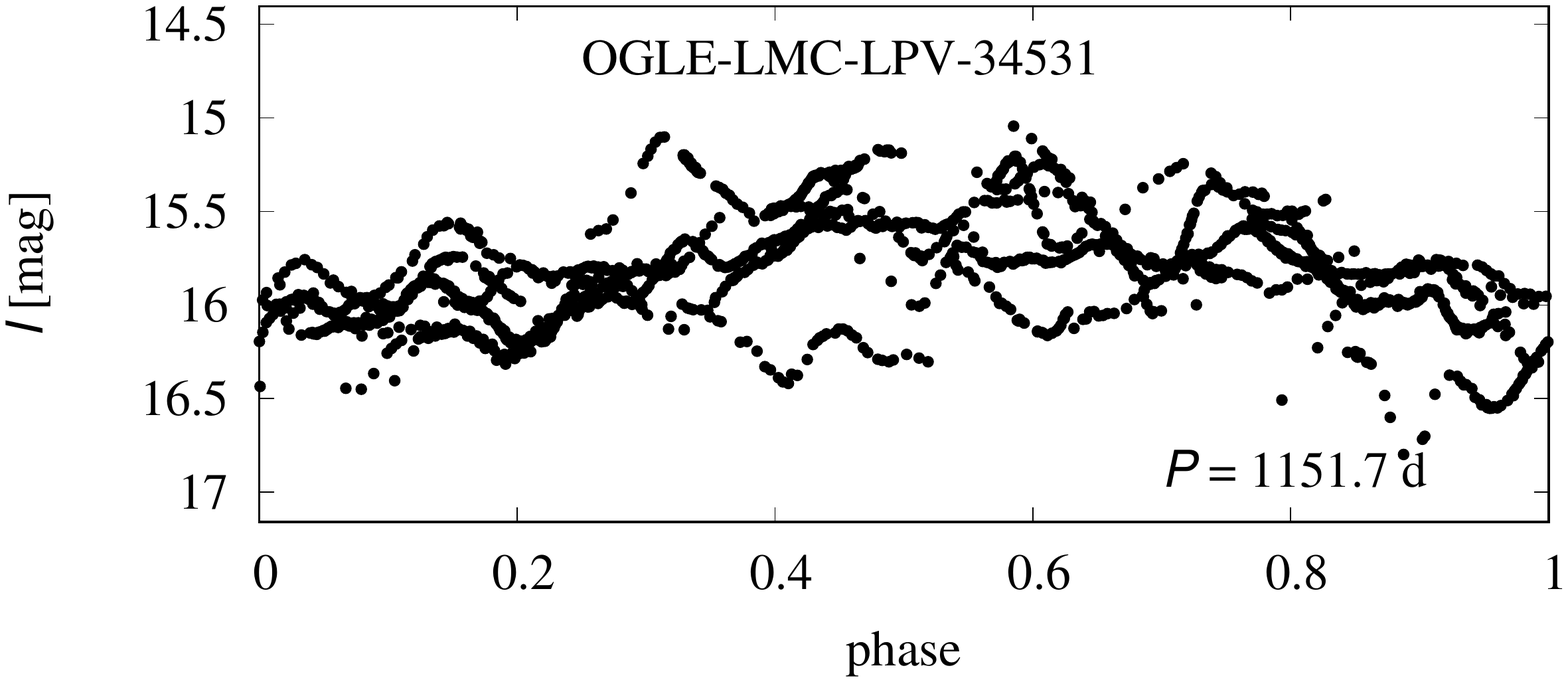}\\
 \includegraphics[width=0.26\textwidth]{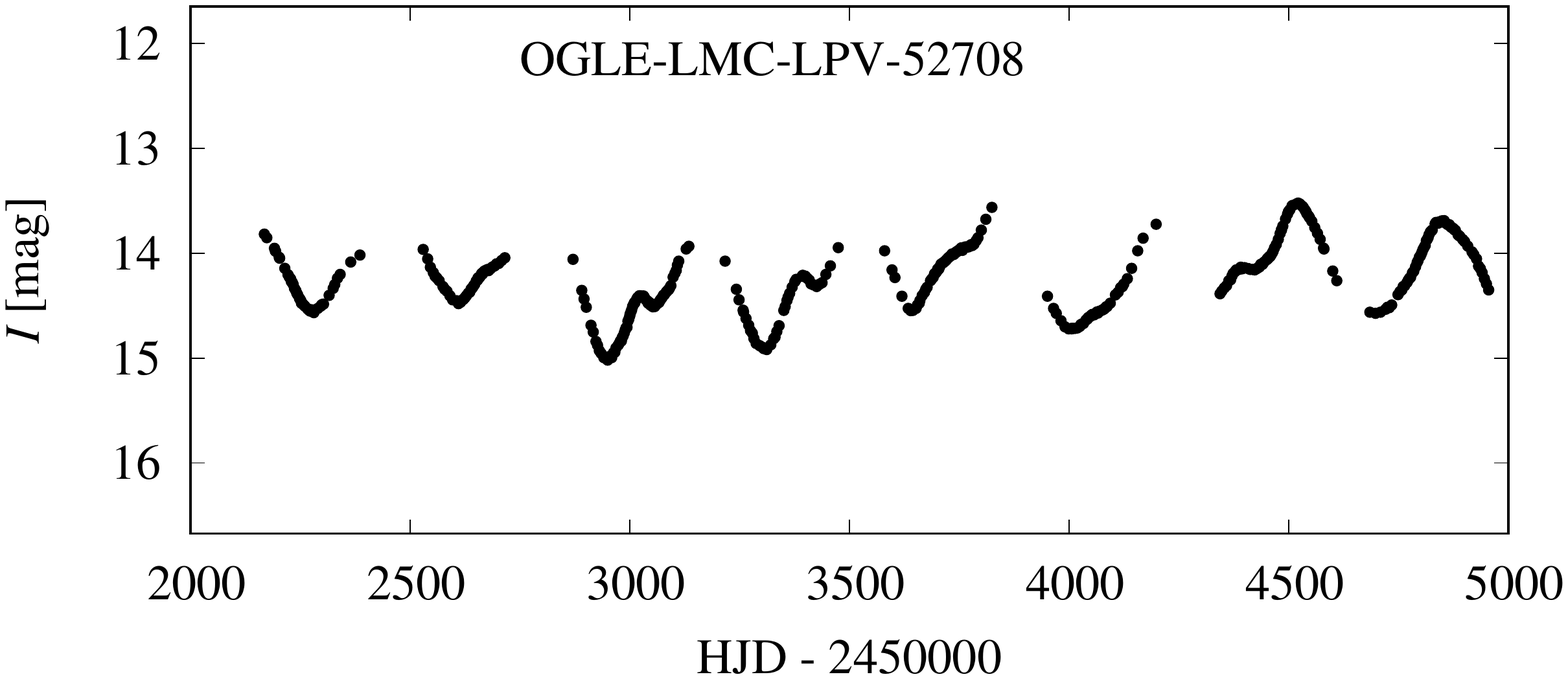} \includegraphics[width=0.26\textwidth]{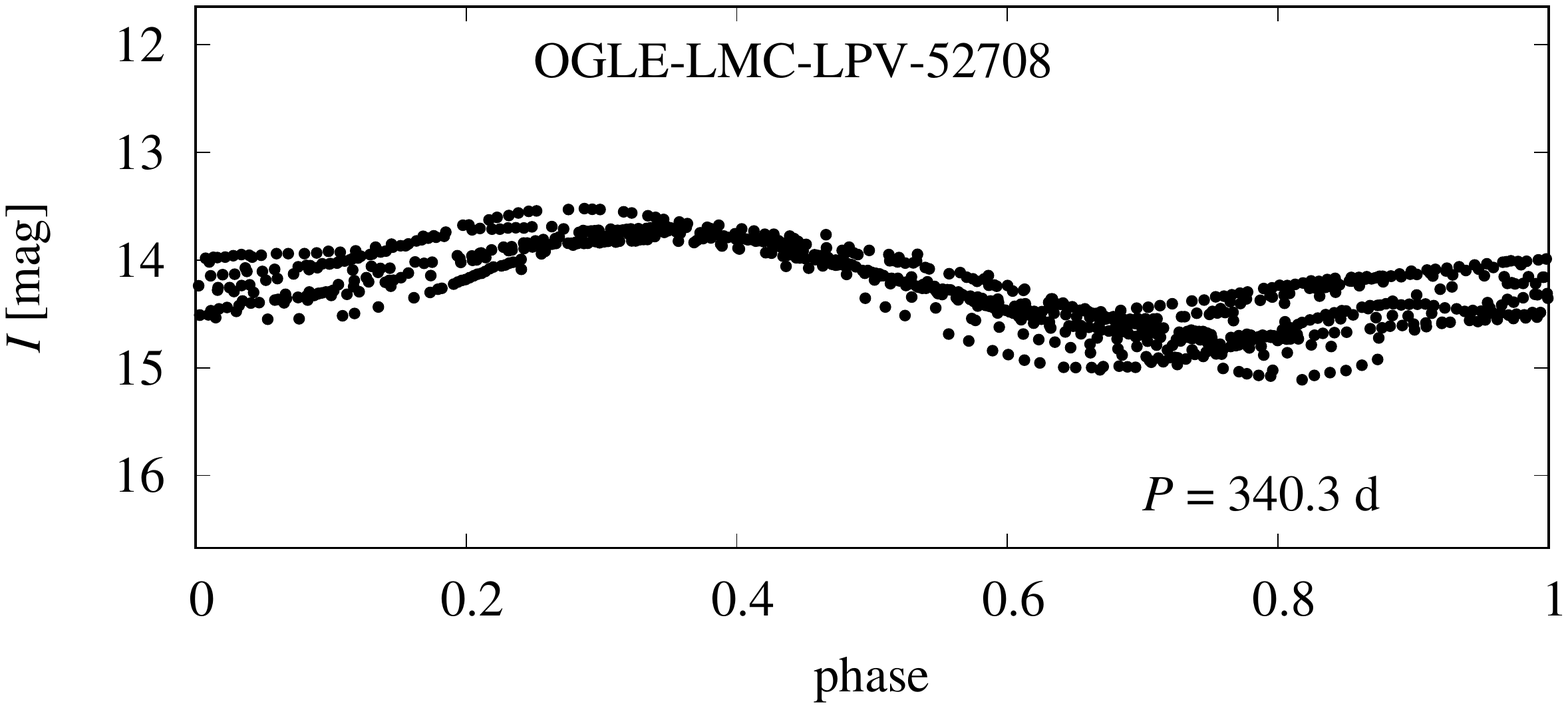}  \includegraphics[width=0.26\textwidth]{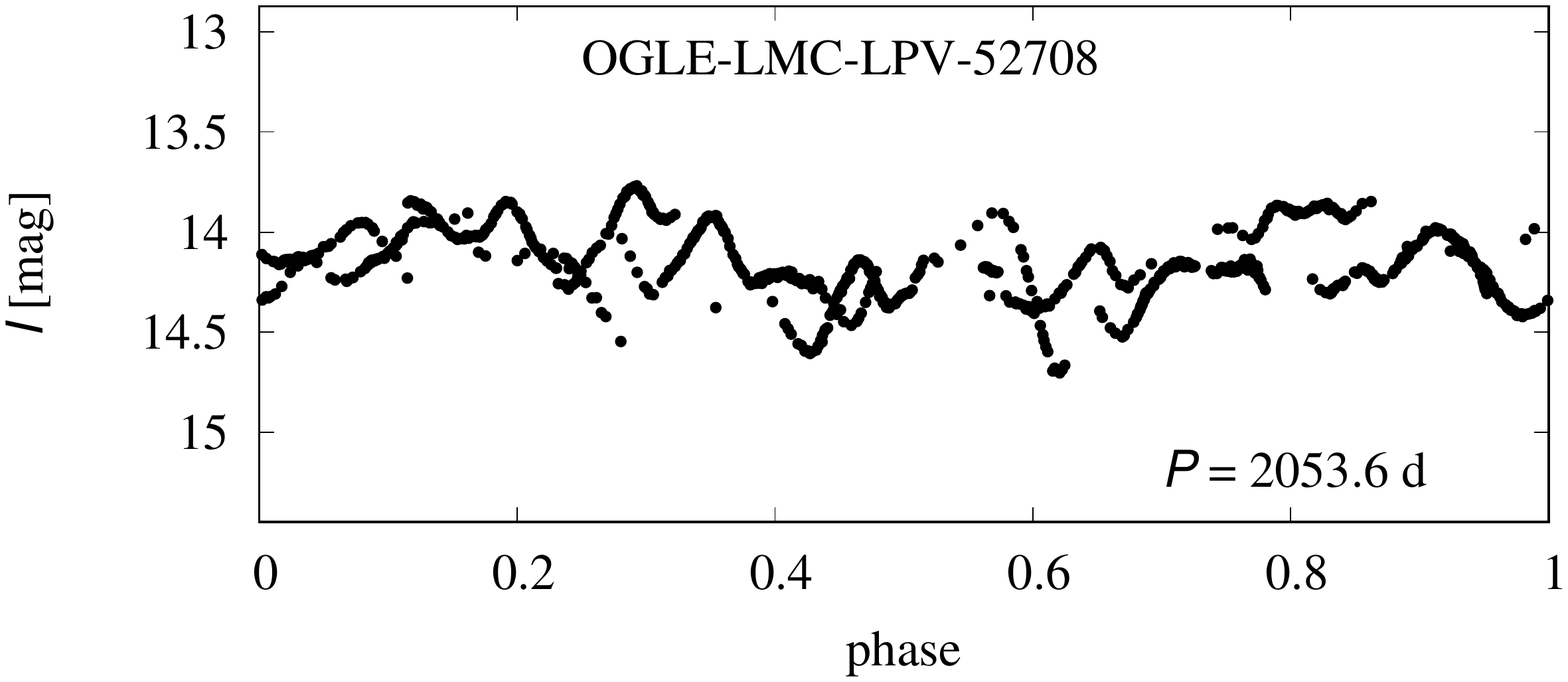}
\caption{Example light curves of the LSP Mira candidates identified in this work. For each star, the original unfolded OGLE light curve is shown in the left panel, the light curve folded with the fundamental mode pulsational period in the middle panel, and the light curve, with the fundamental mode prewhitened, folded with the putative LSP in the right panel.}
\end{figure*}

\begin{figure*}
\label{lcs}
 \includegraphics[width=0.26\textwidth]{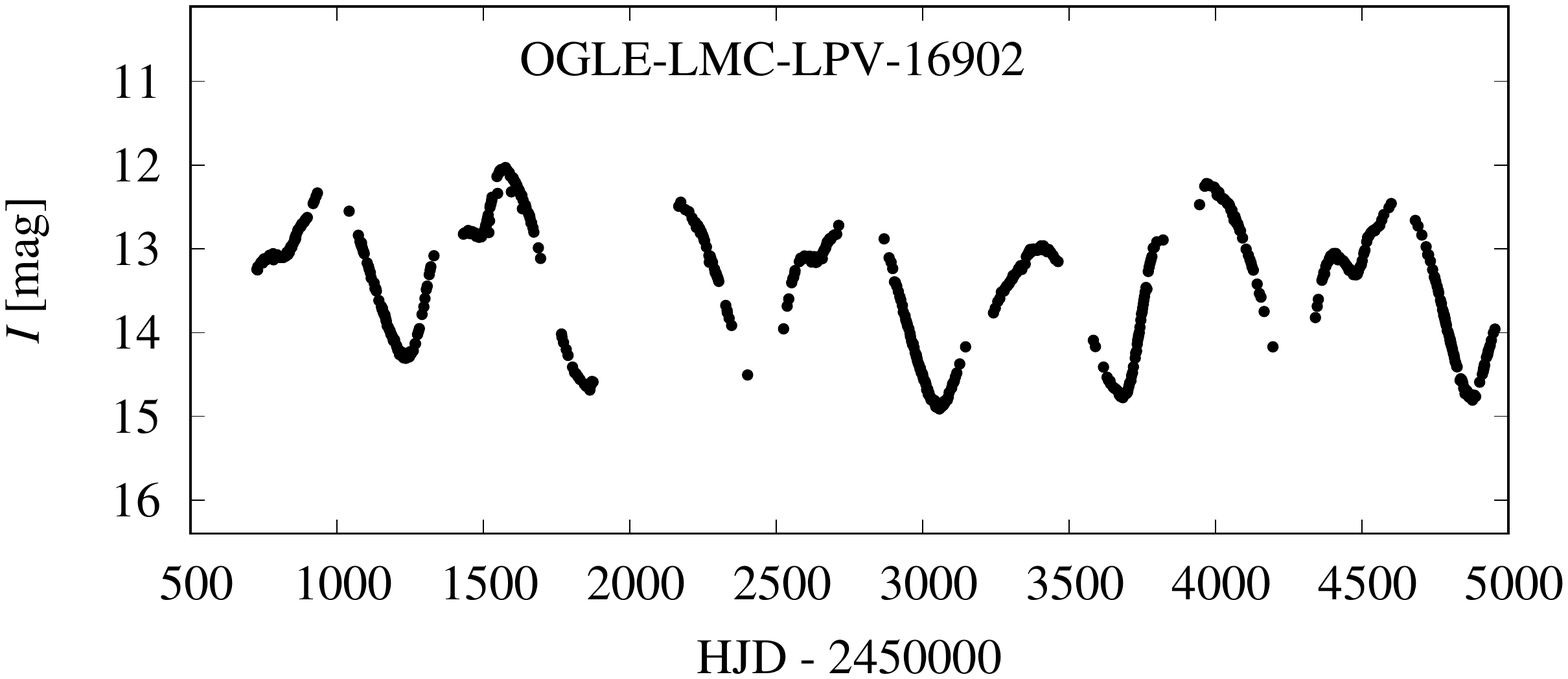} \includegraphics[width=0.26\textwidth]{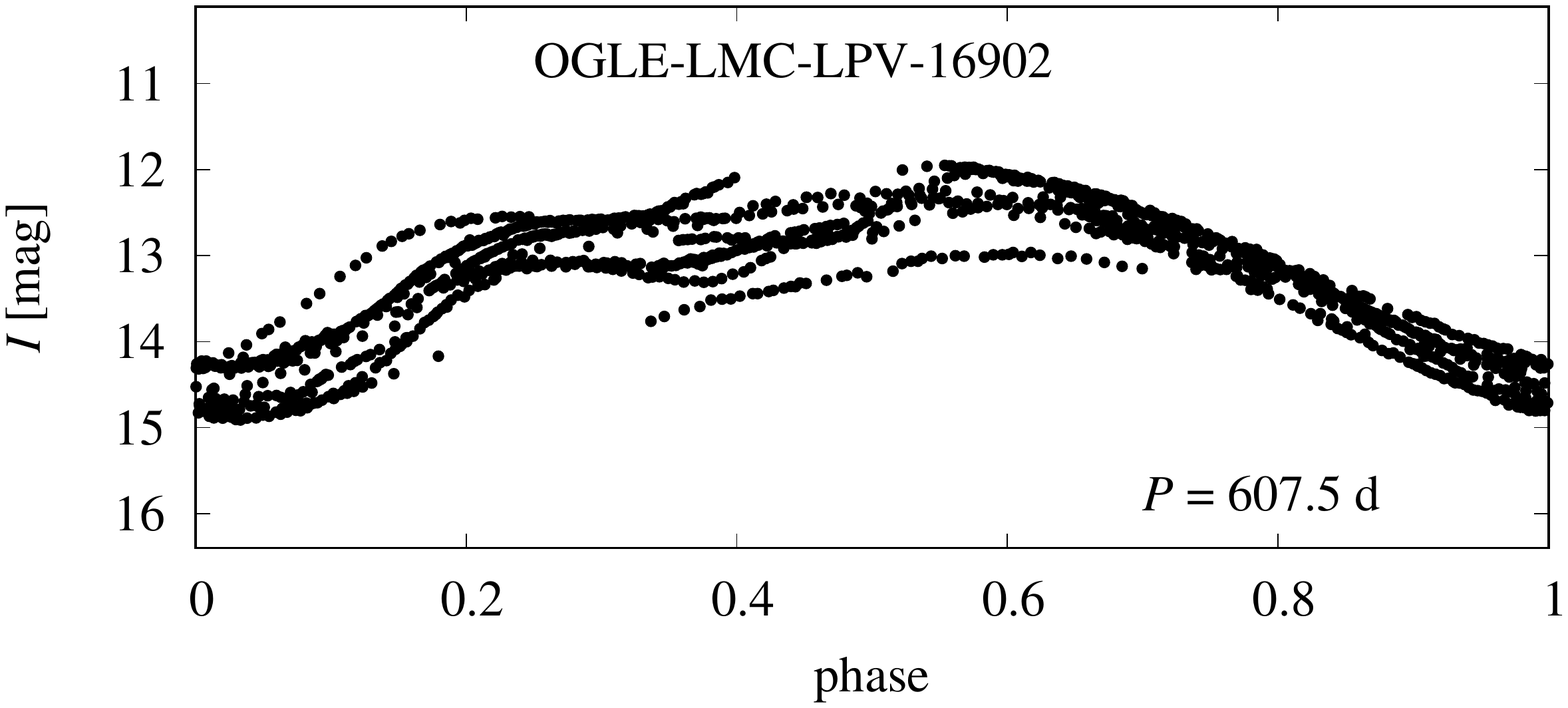}  \includegraphics[width=0.26\textwidth]{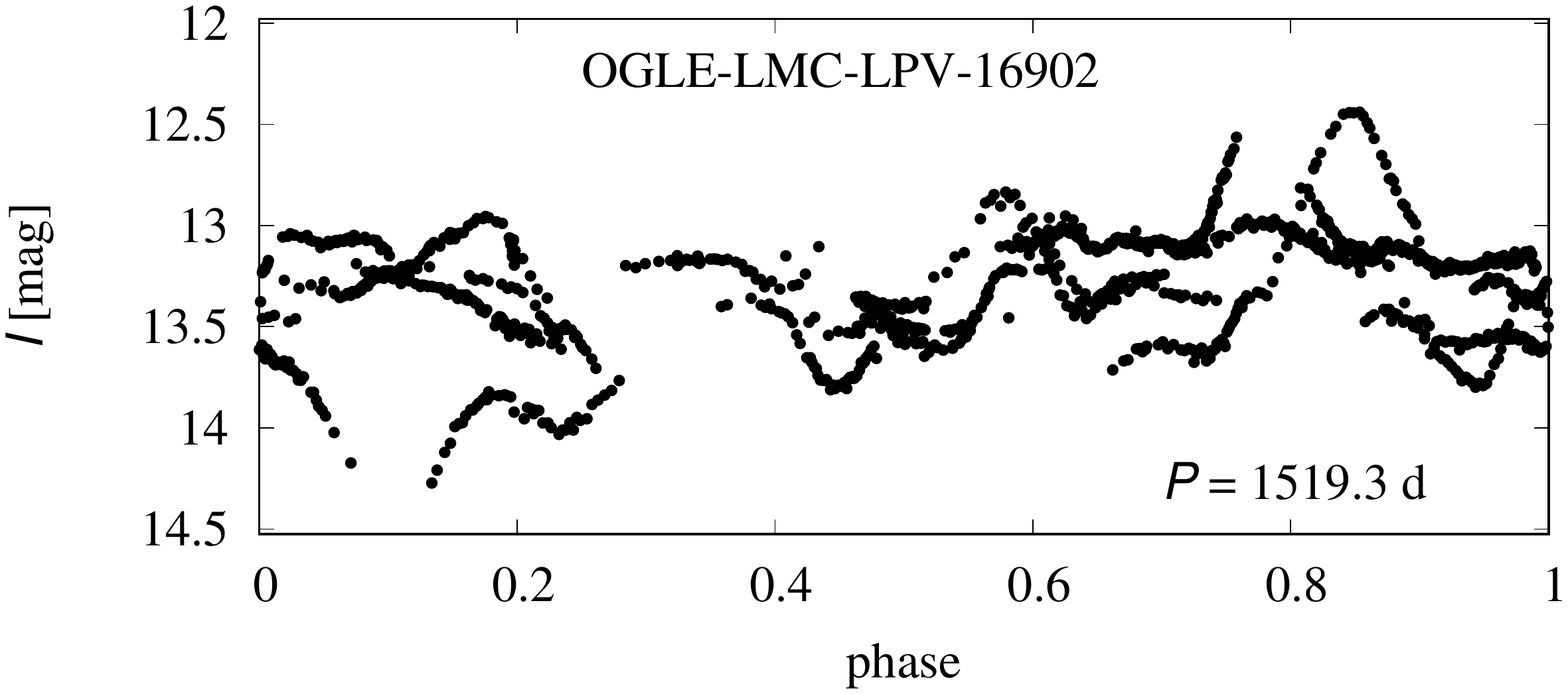} \\
 \includegraphics[width=0.26\textwidth]{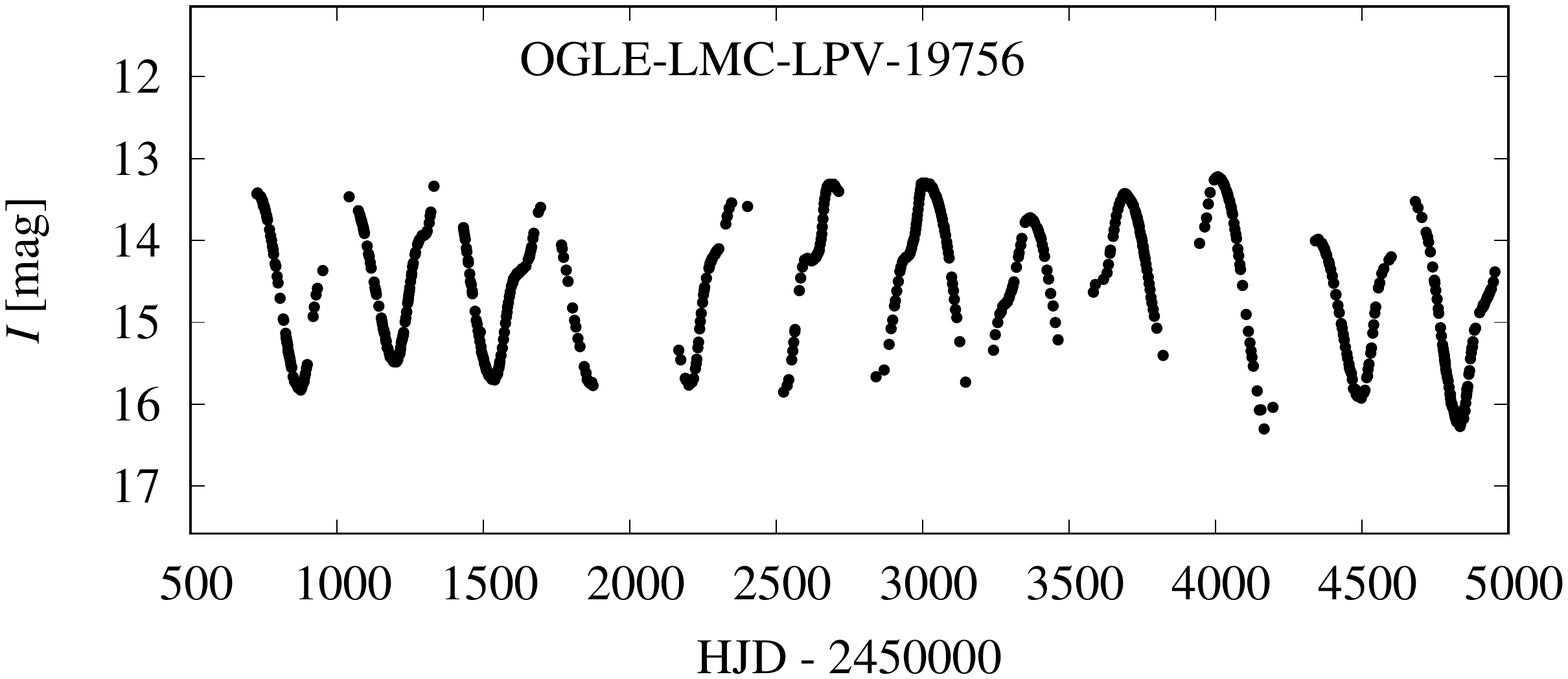} \includegraphics[width=0.26\textwidth]{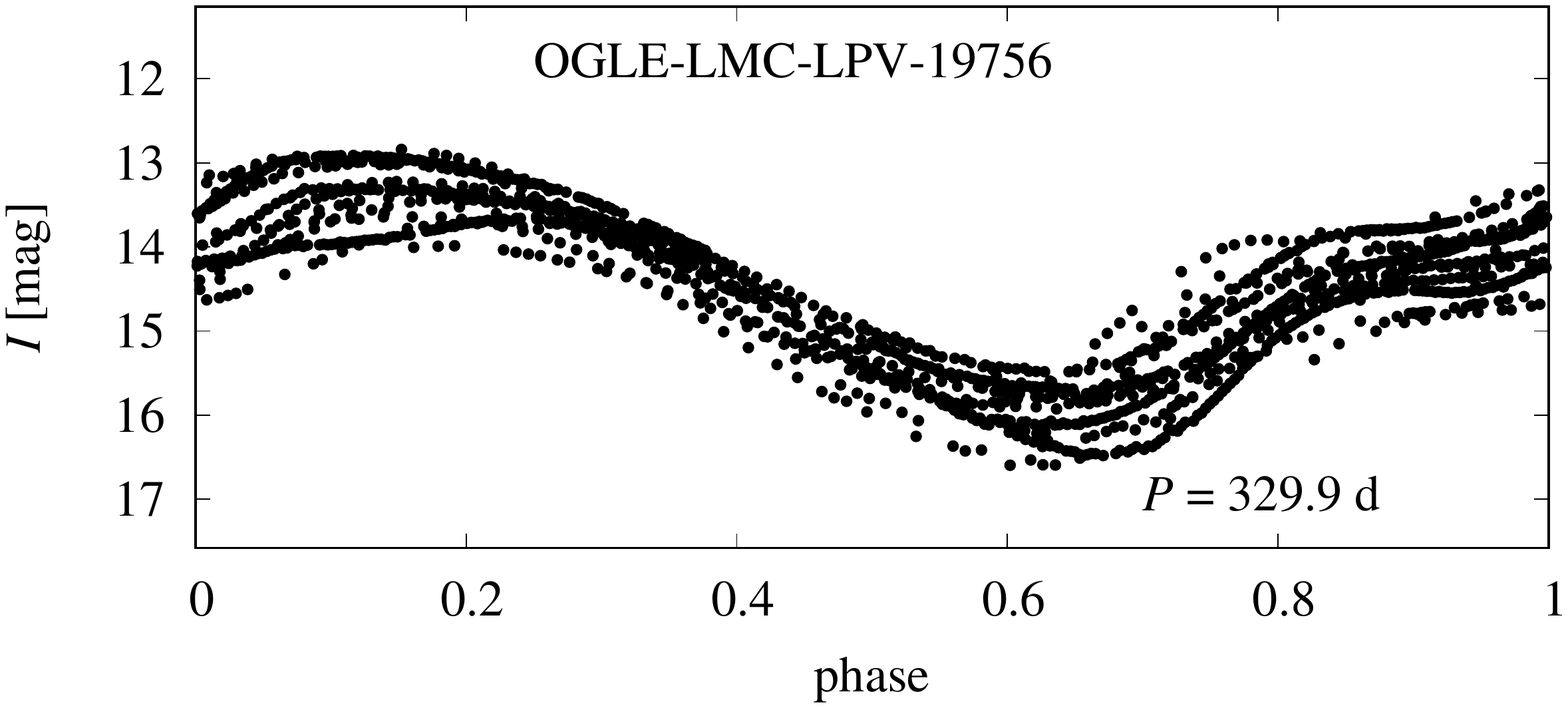}  \includegraphics[width=0.26\textwidth]{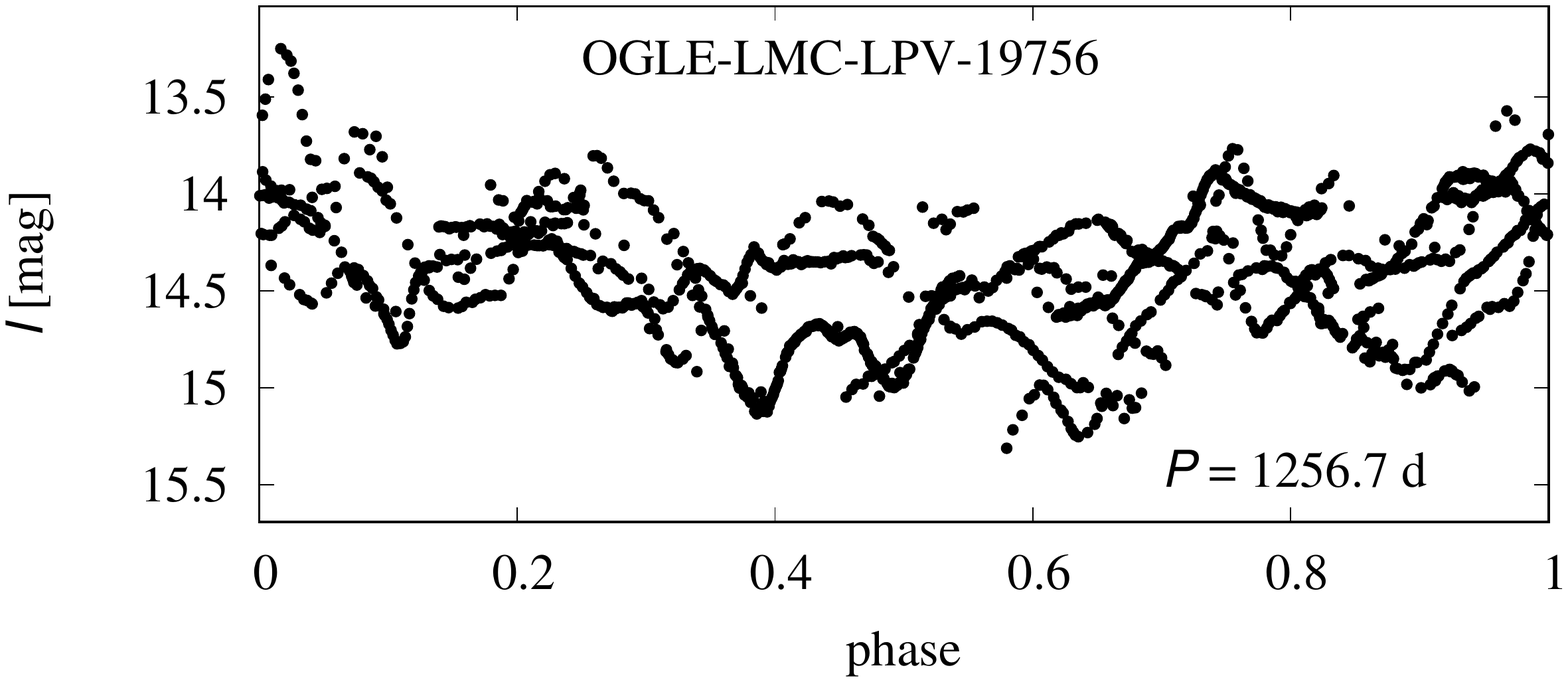} \\
 \includegraphics[width=0.26\textwidth]{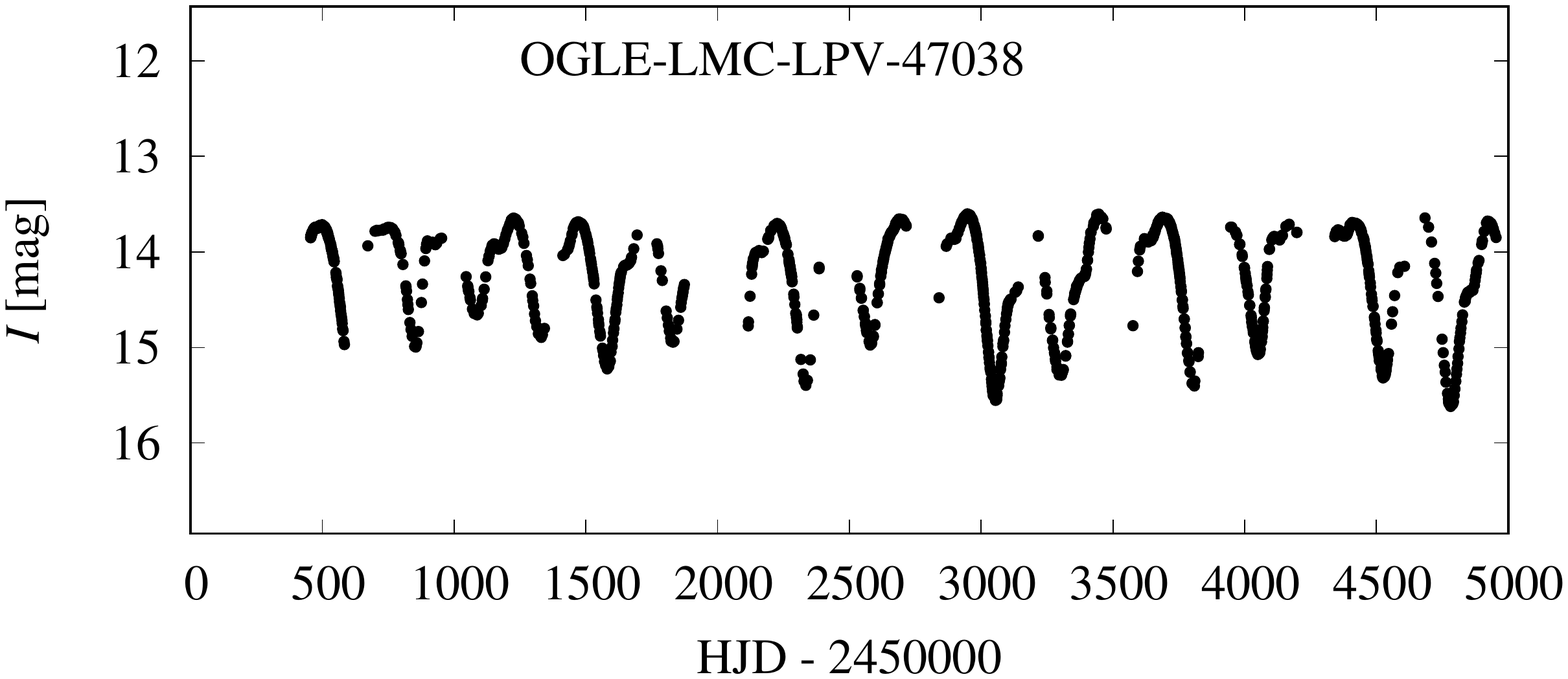} \includegraphics[width=0.26\textwidth]{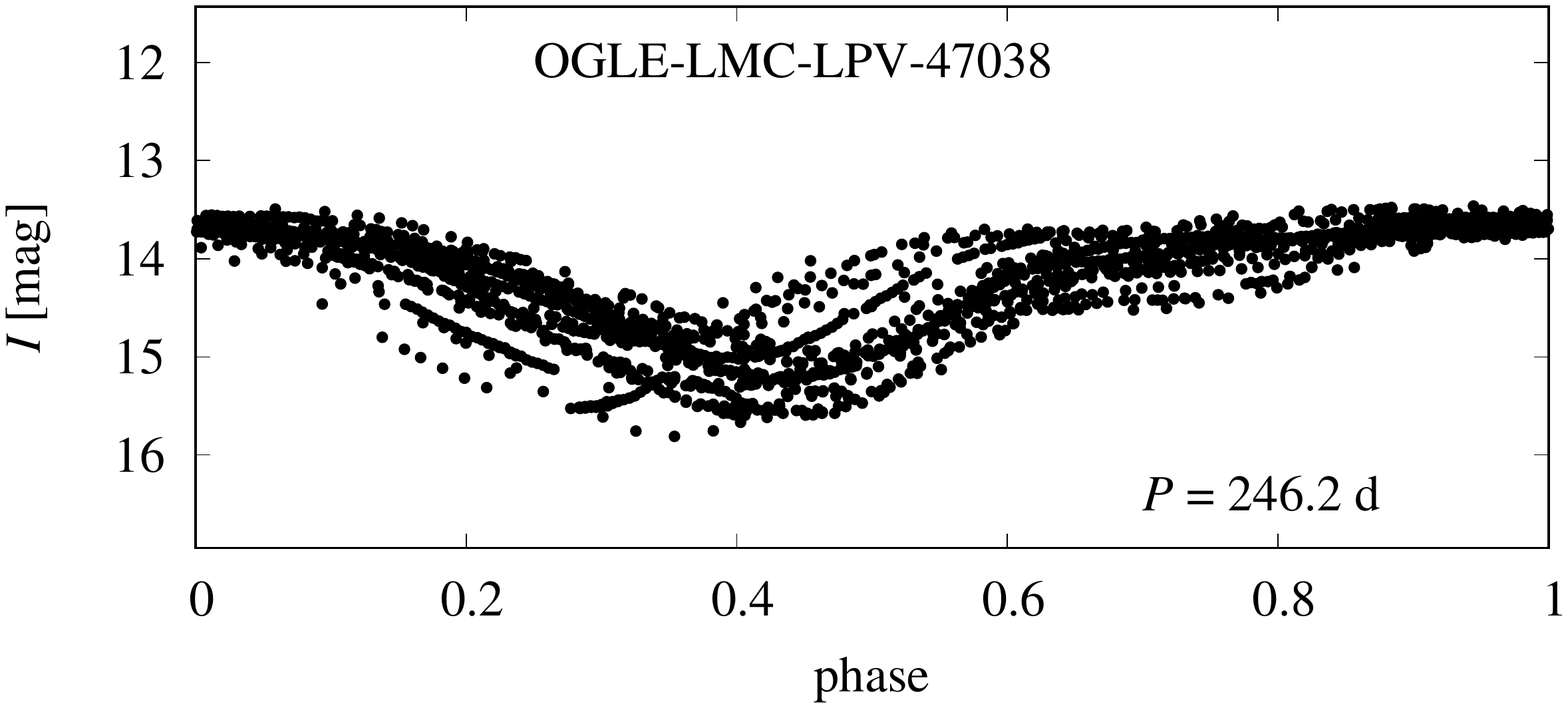}  \includegraphics[width=0.26\textwidth]{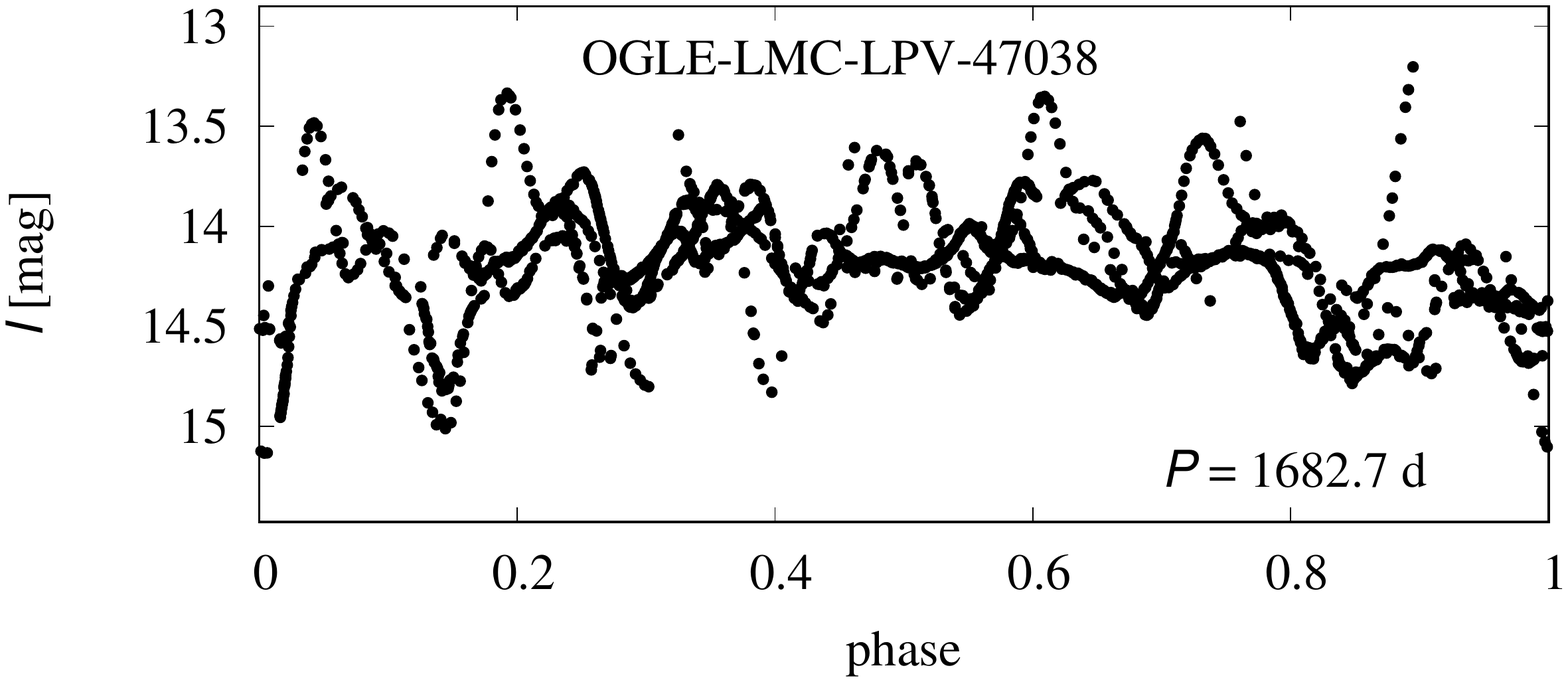}\\
 \includegraphics[width=0.26\textwidth]{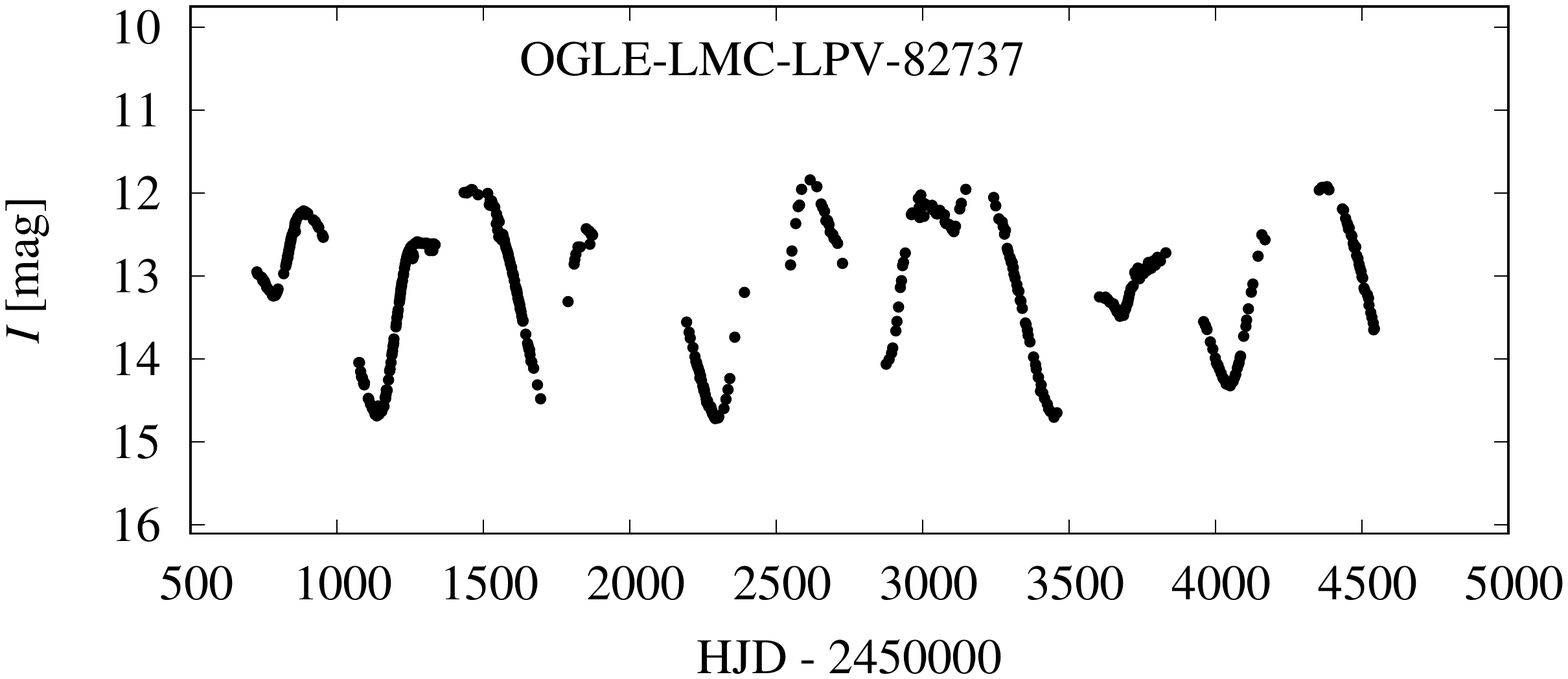} \includegraphics[width=0.26\textwidth]{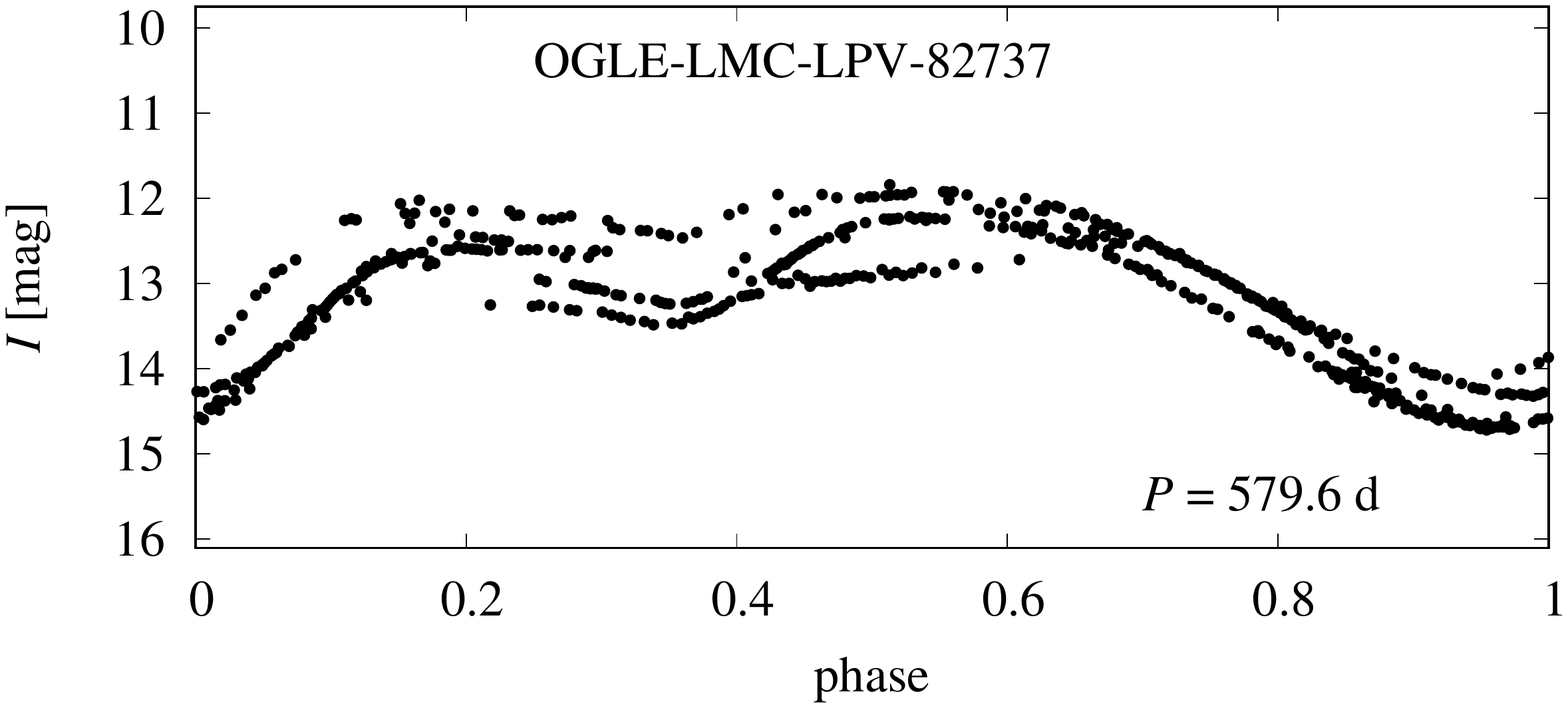}  \includegraphics[width=0.26\textwidth]{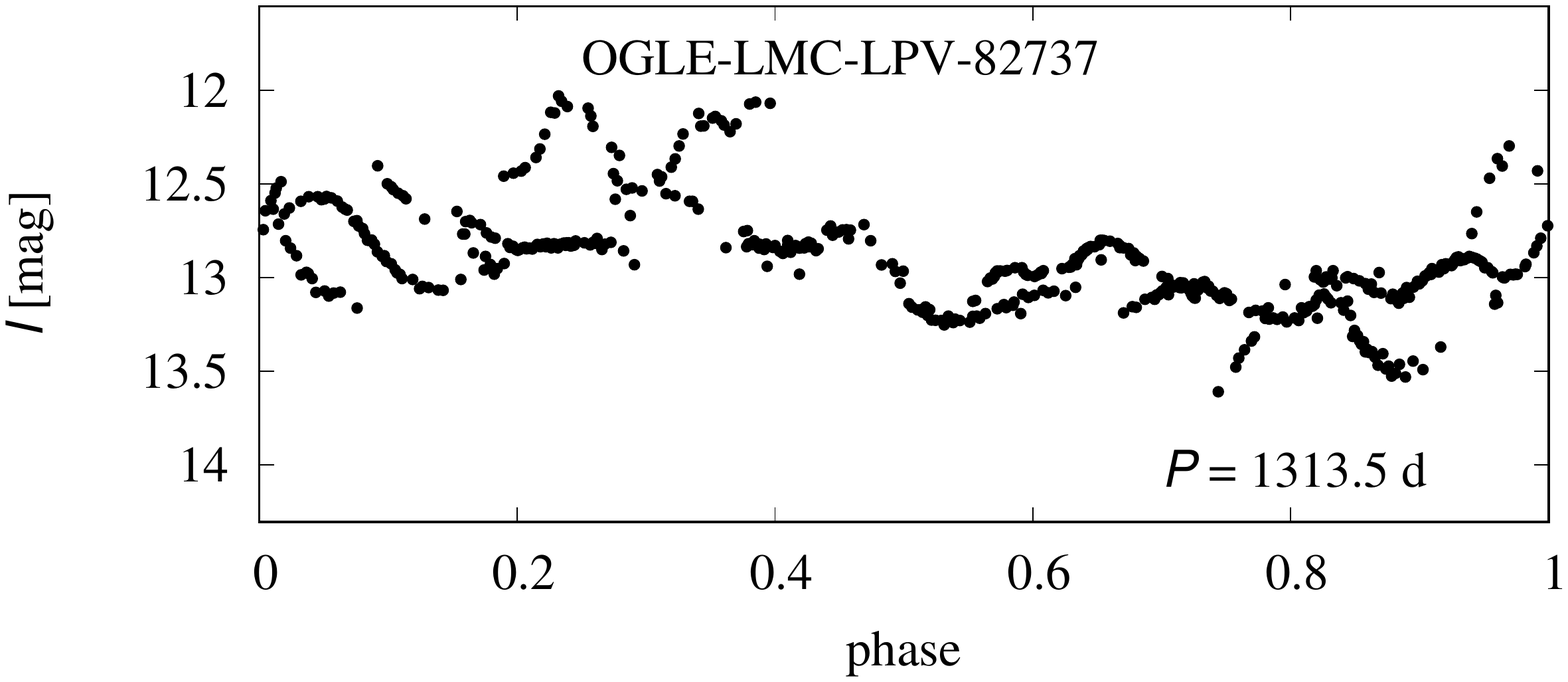}\\
 \includegraphics[width=0.26\textwidth]{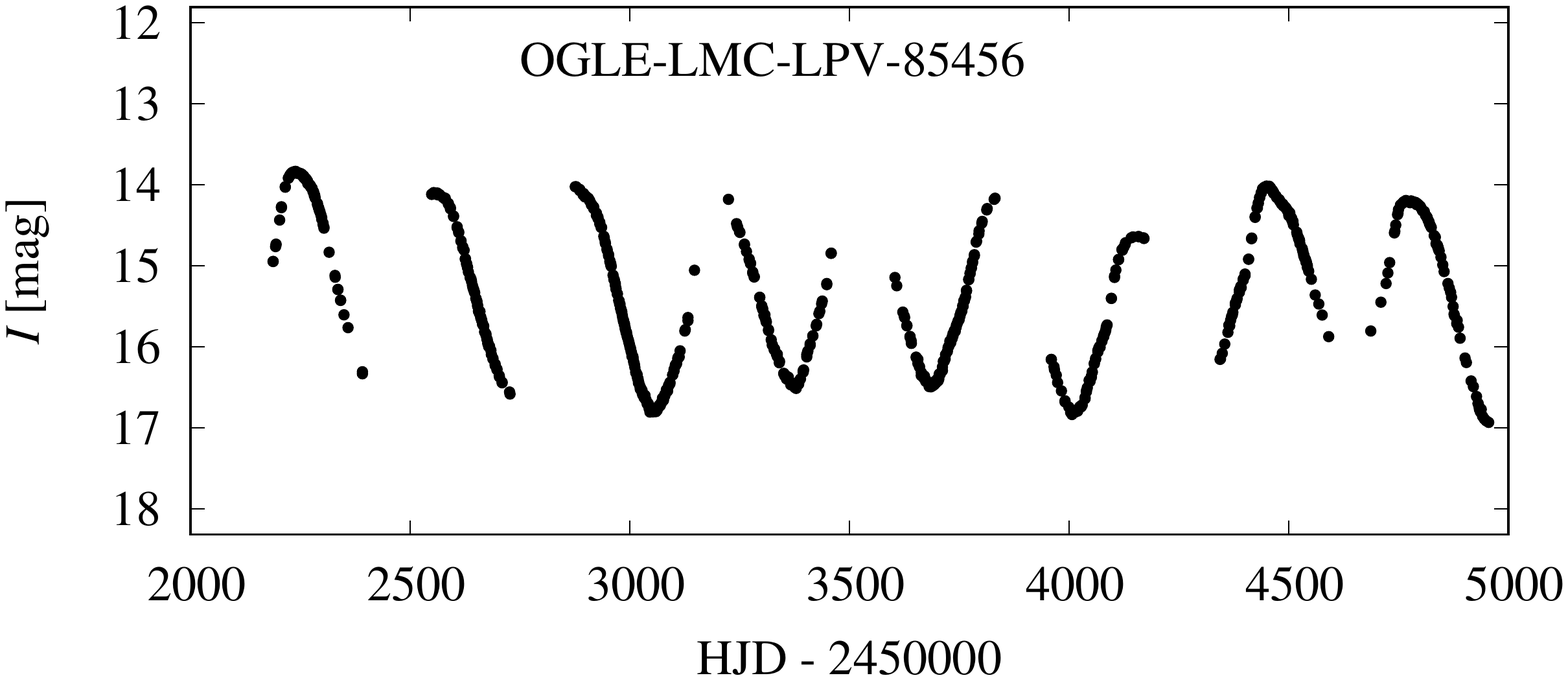} \includegraphics[width=0.26\textwidth]{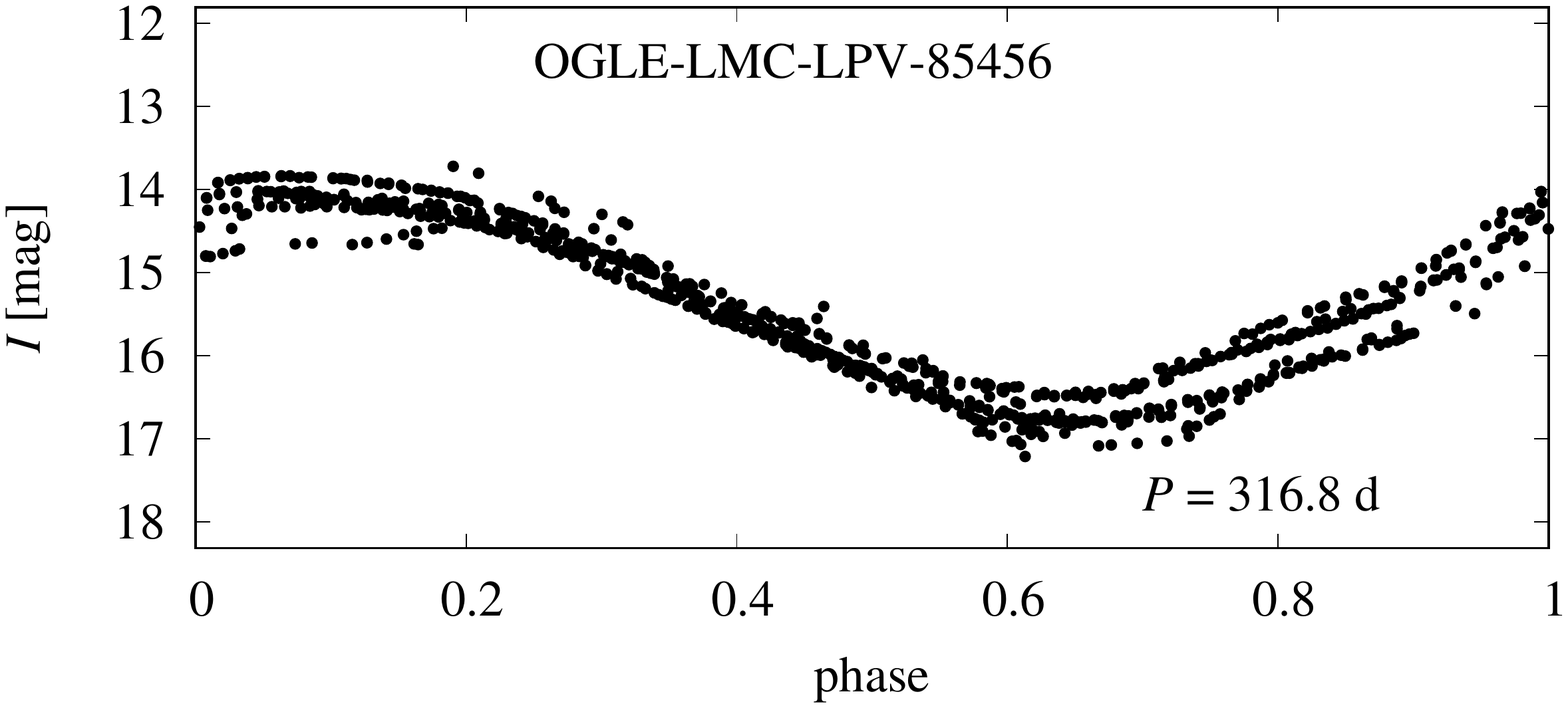}  \includegraphics[width=0.26\textwidth]{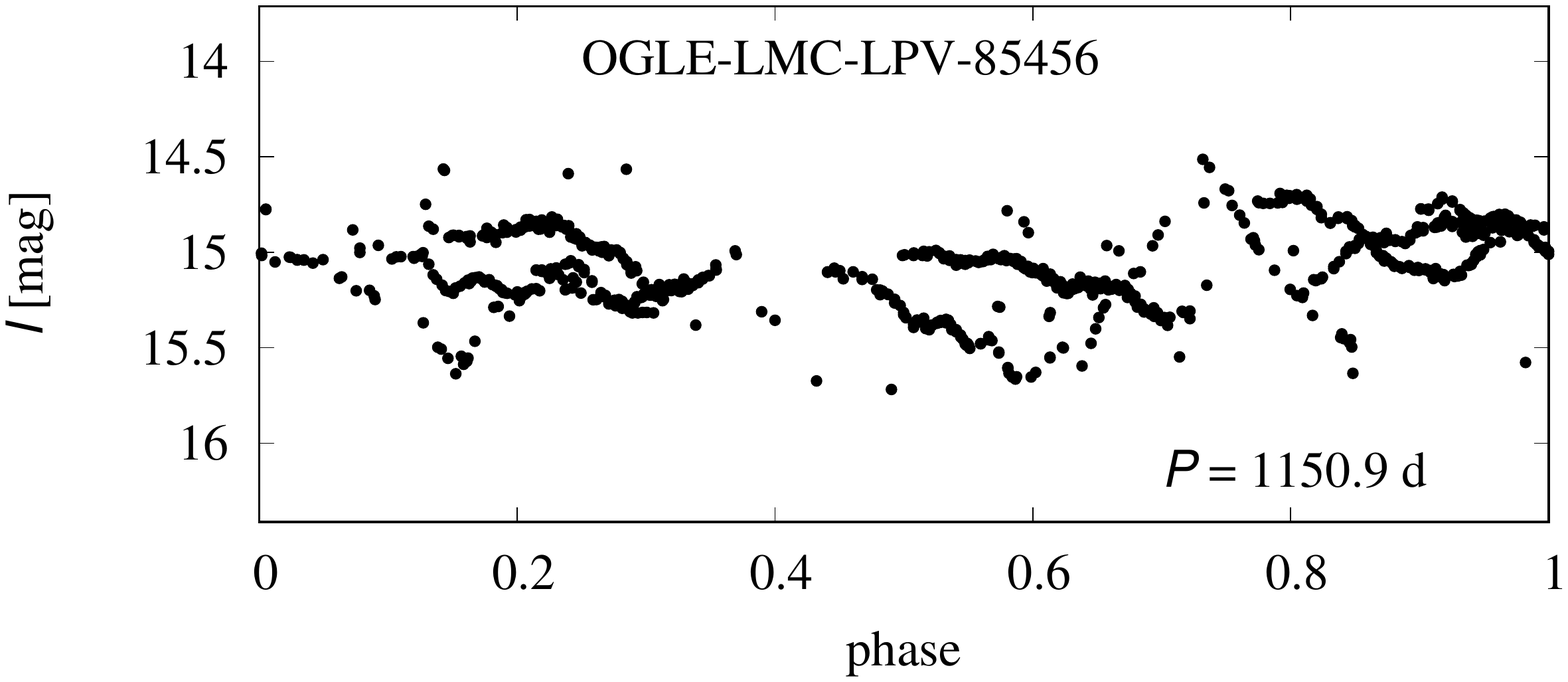}\\
 \includegraphics[width=0.26\textwidth]{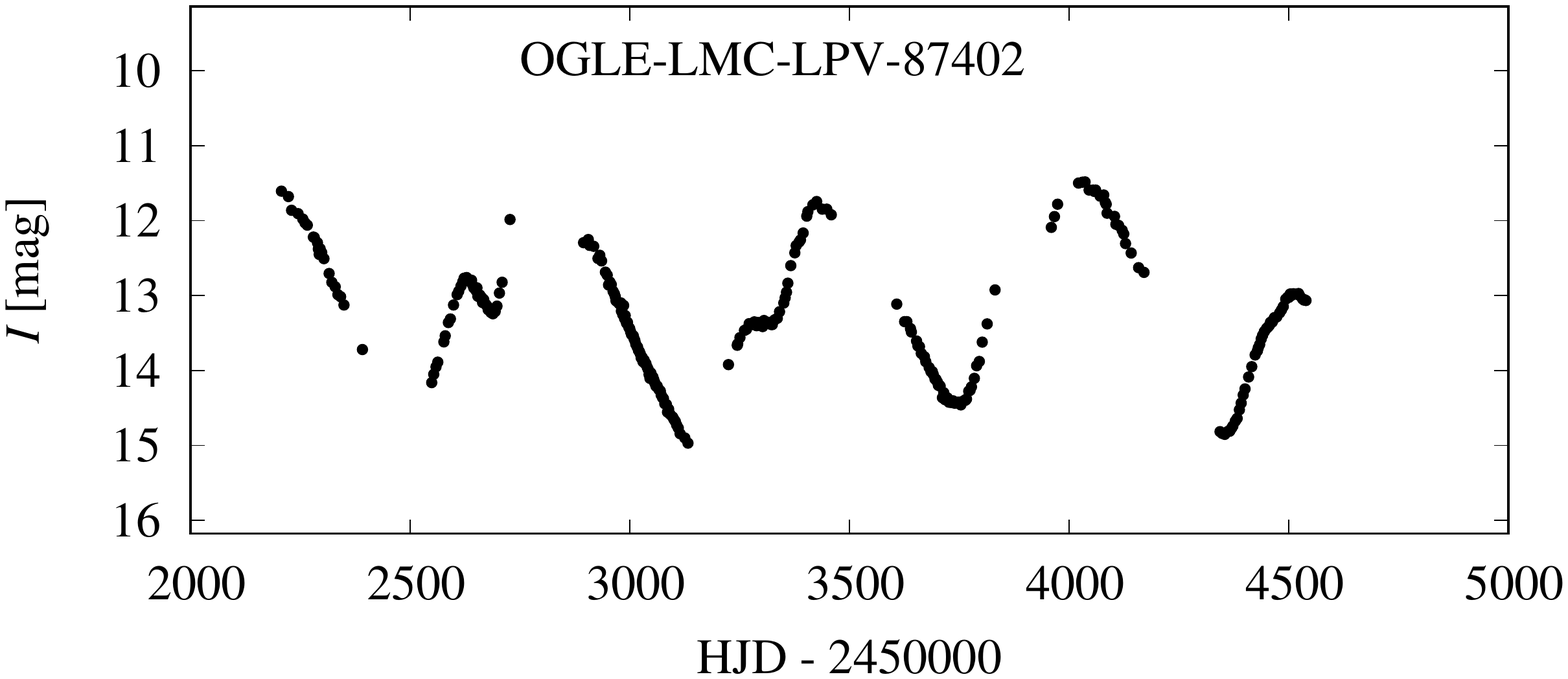} \includegraphics[width=0.26\textwidth]{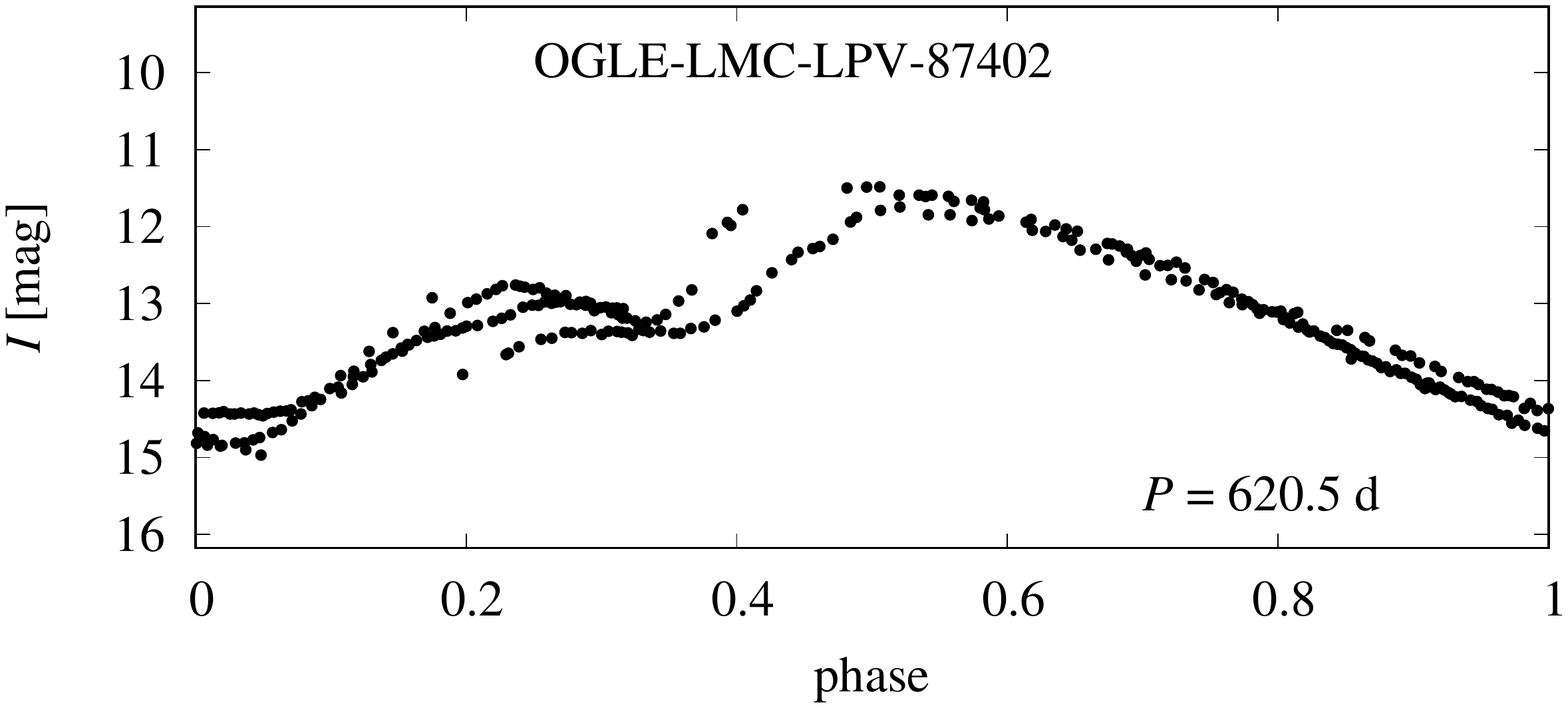}  \includegraphics[width=0.26\textwidth]{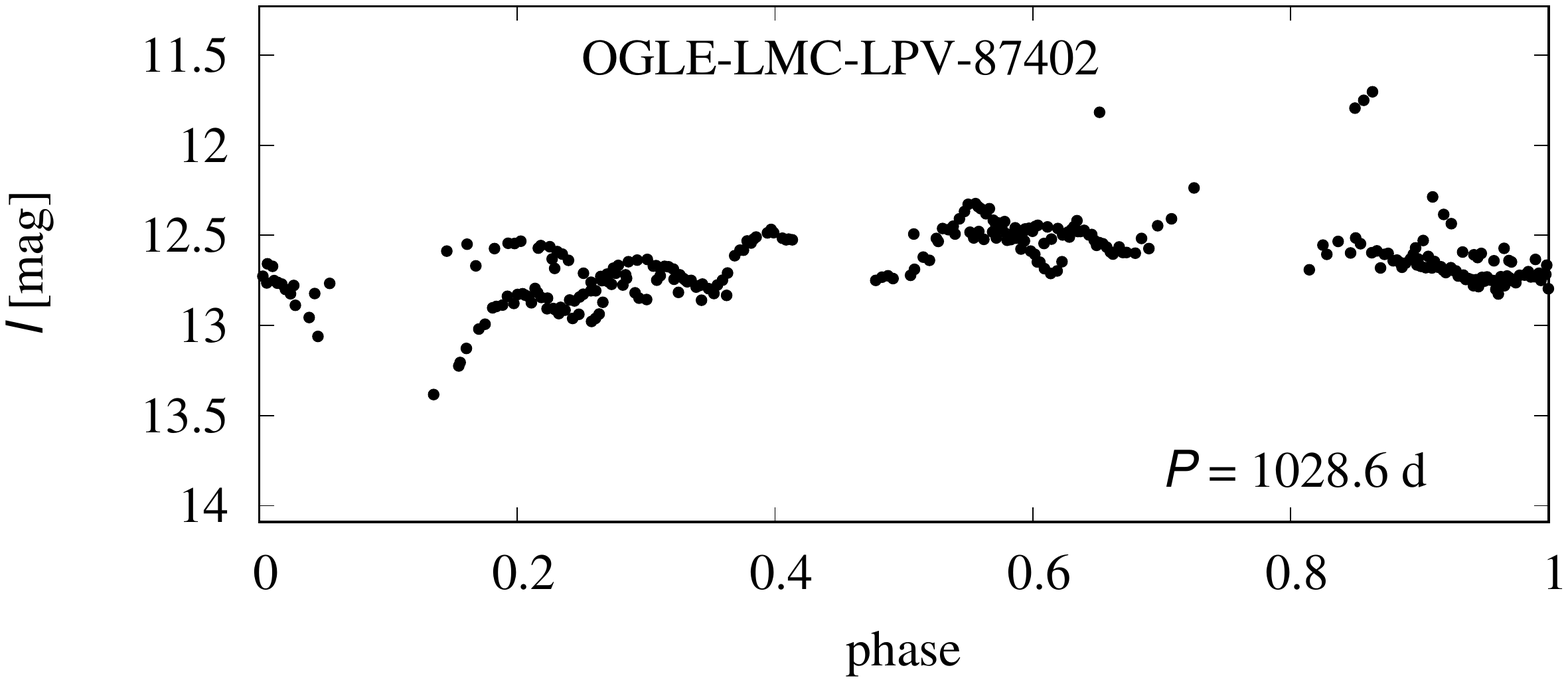}\\
 \includegraphics[width=0.26\textwidth]{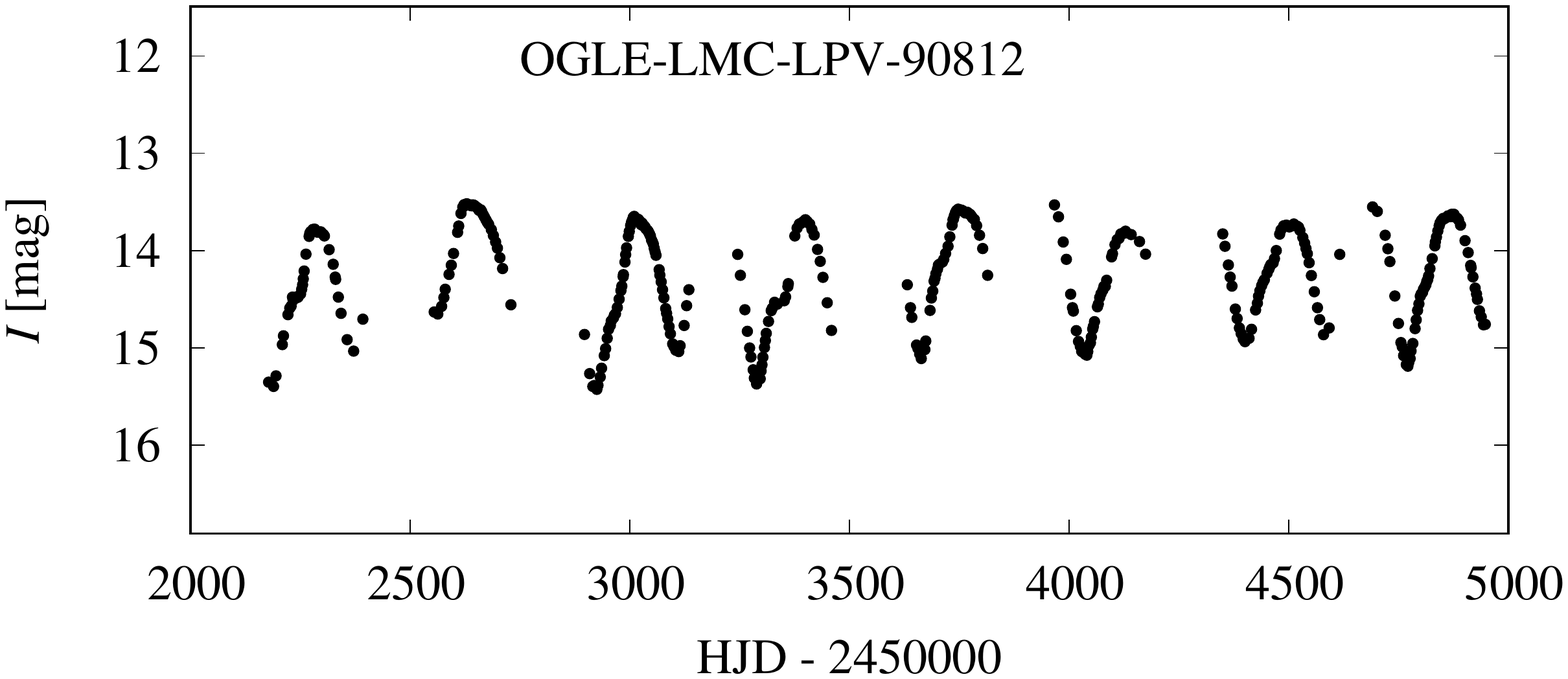} \includegraphics[width=0.26\textwidth]{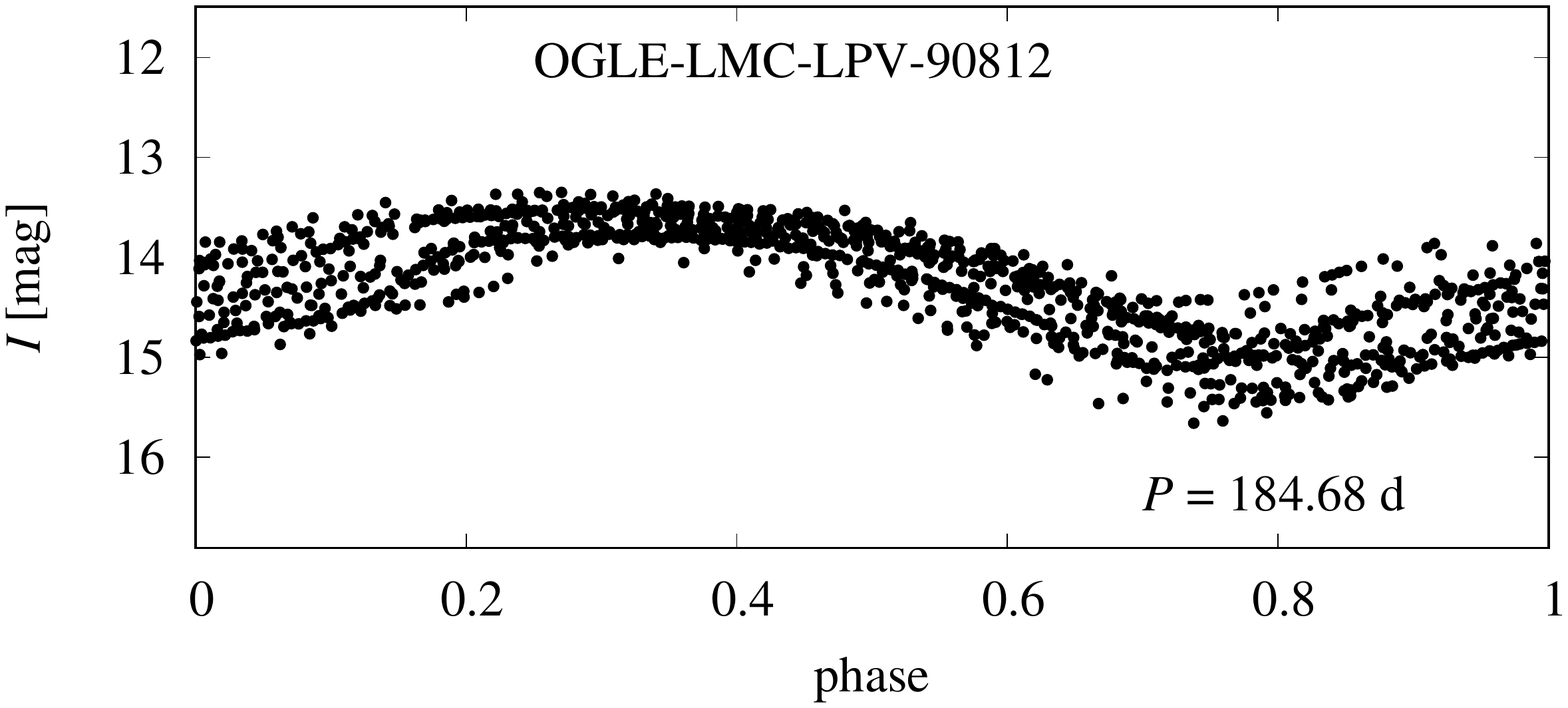}  \includegraphics[width=0.26\textwidth]{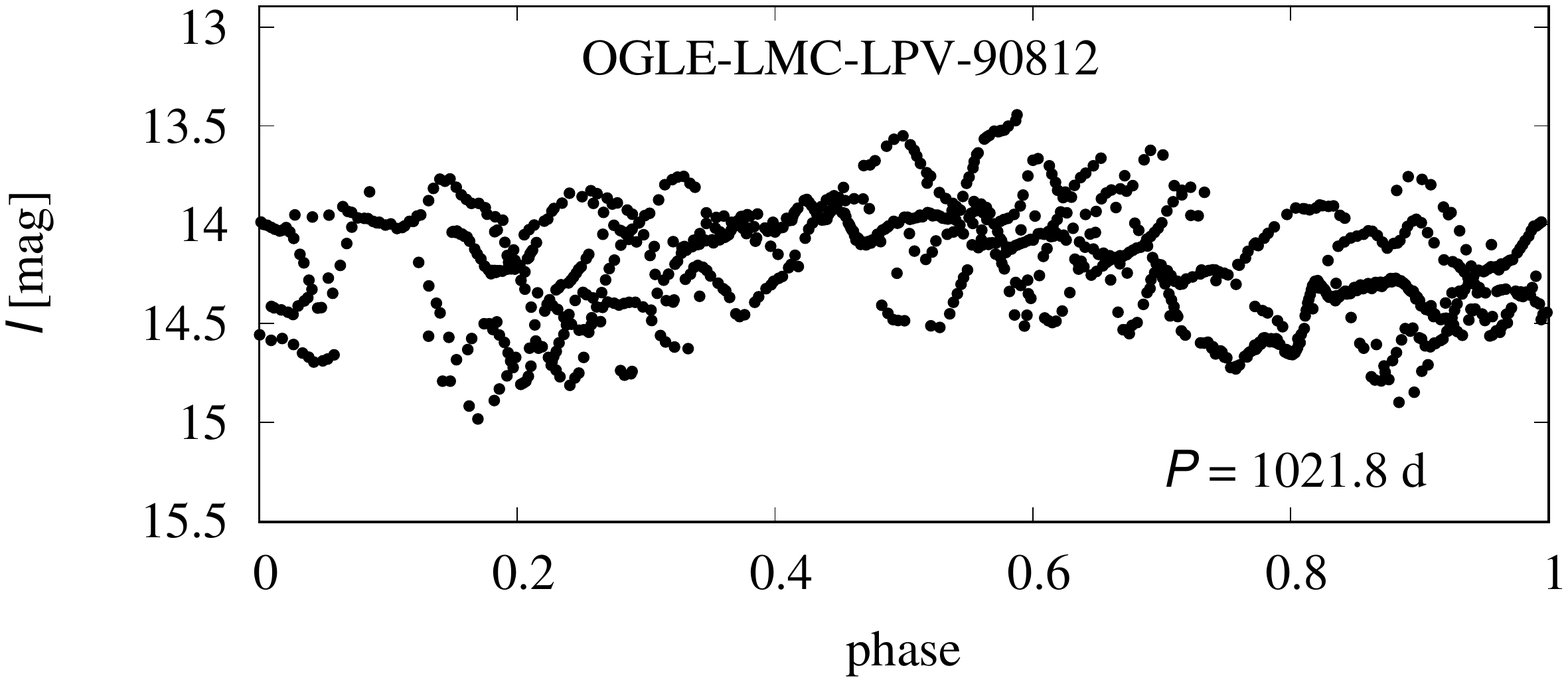}
  
\caption{The same as Fig.~3, but for O-rich candidates only.}
\end{figure*}

\begin{figure*}[ht]
\centering
	\includegraphics[width=0.99\textwidth]{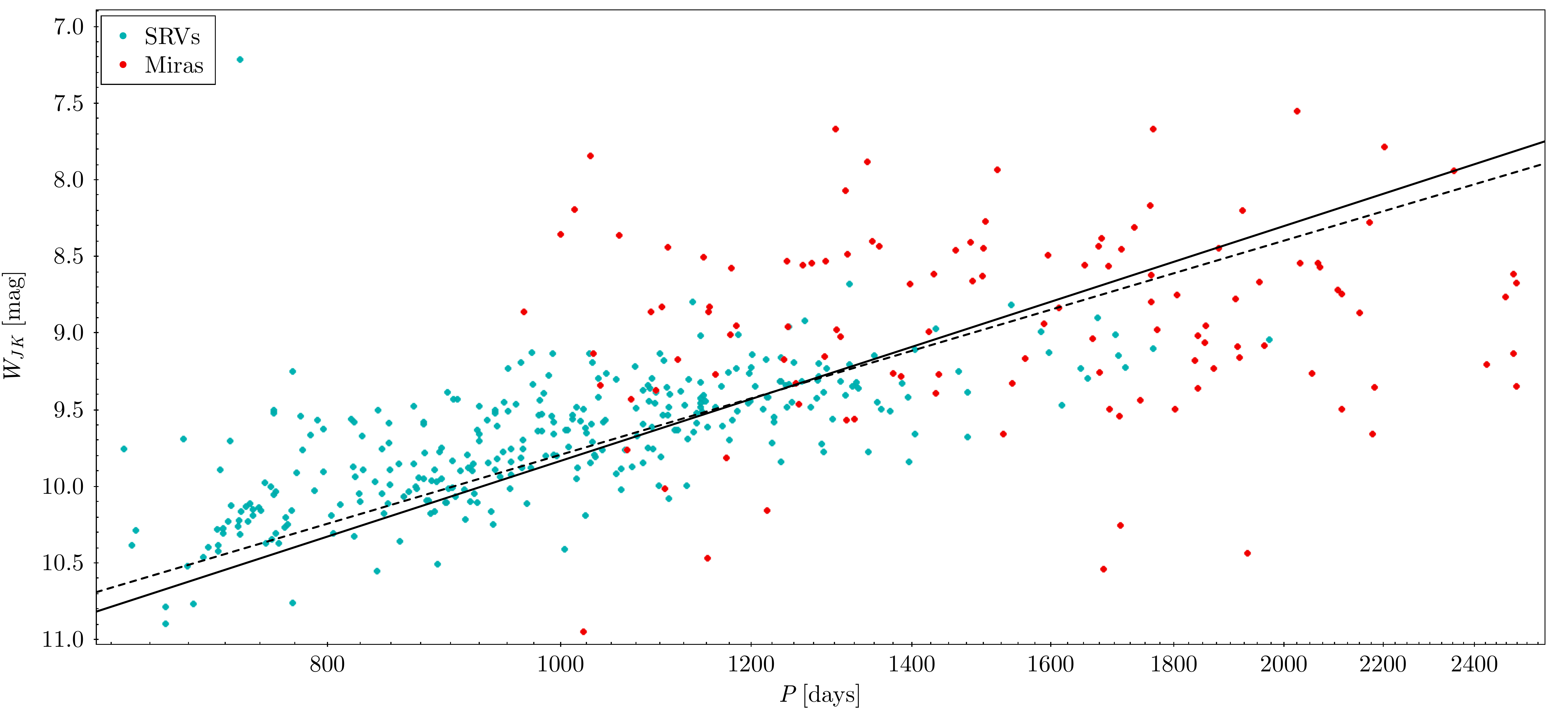} 
	\caption{ PLR in $W_{JK}$ index formed by Mira LSP candidates. C-rich Miras are known LSP SRVs. PLR derived for LSPs by \citet{soszynskietal2007} are marked with solid lines (C-rich - solid, O-rich - dashed line).  }  
	    \label{fig:mira_srv}
\end{figure*}

\begin{figure*}[ht]
\centering
	\includegraphics[width=0.99\textwidth]{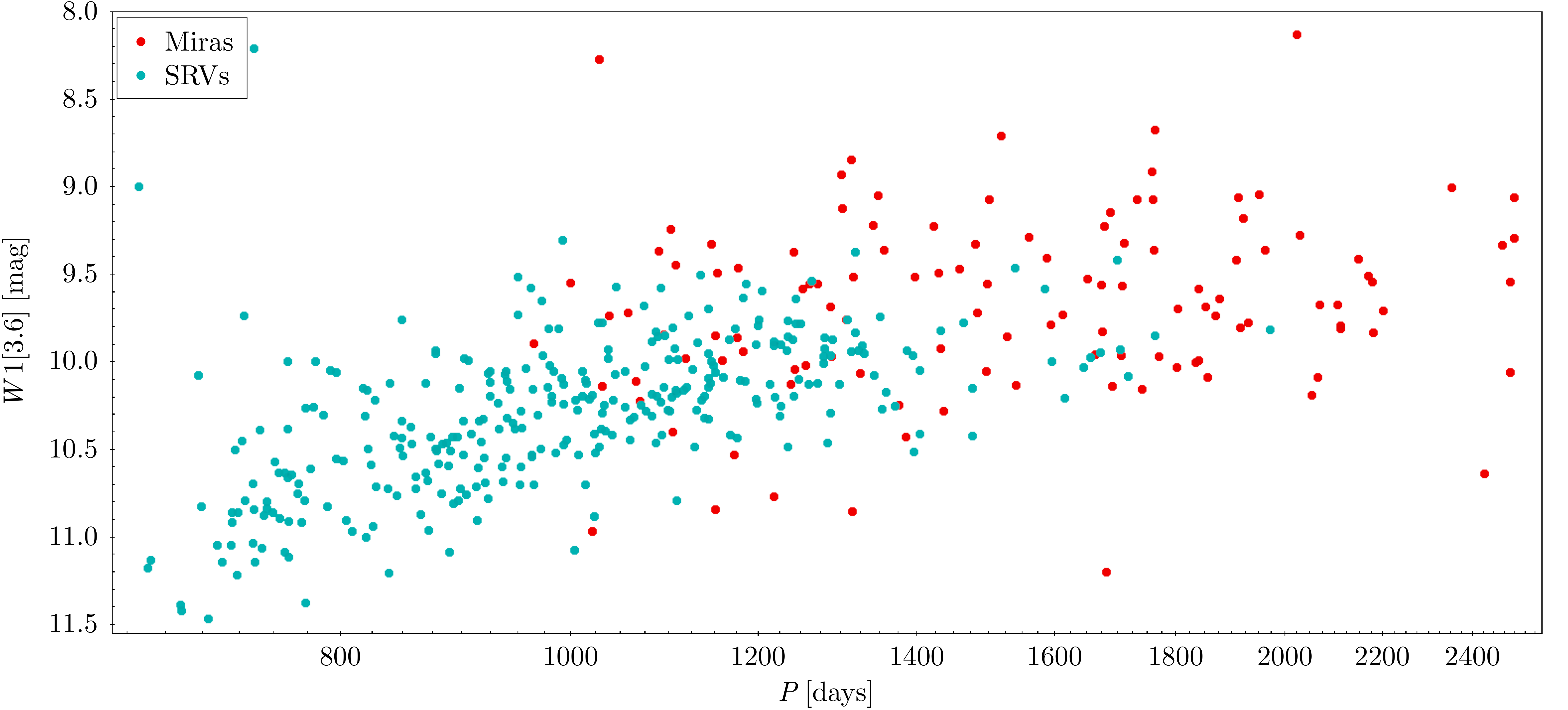} 
	\caption{PLR in the Wise [3.6] band.}  
	    \label{fig:ampl}
\end{figure*}

\begin{figure*}[ht]
\centering
	\includegraphics[width=0.99\textwidth]{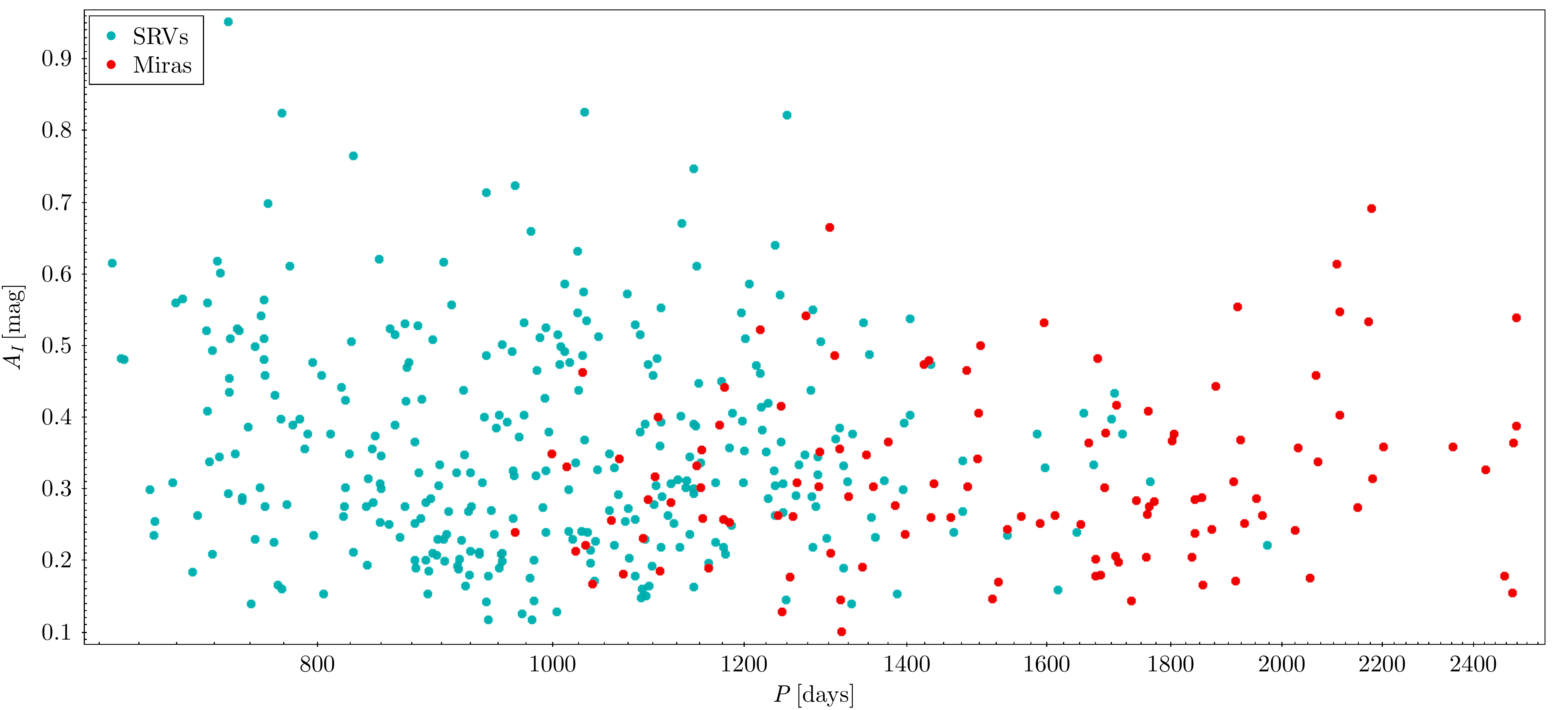} 
	\caption{ Amplitudes comparison for LSP SRVs and candidates for LSP Miras.}  
	    \label{fig:ampl}
\end{figure*}

For each of the Miras, I subtracted the main variability from the $I$-band light curve, using the {\it killharm} routine included in the VARTOOLS data analysis package \citep{hartman2016}. For this, I used the strongest period given by OGLE and subtracted a four-harmonic model. Miras variability is not as strictly periodic as for example Classical Cepheids, therefore this procedure does not remove the original periodicity completely. However, it does remove the dominant peak, corresponding to it, from the Fourier frequency spectrum. This allowed for automatic detection of the secondary peak if present. The procedure also dumped the original variability to the level that allowed for visual inspection of secondary periodicity, which would have been otherwise obscured by the primary, high-amplitude pulsation. 

I run a new period search on each of the period-subtracted light curves with the Generalized Lomb-Scargle periodogram method \citep{zechmeister2009} using the {\it LS} routine from VARTOOLS package. First, I attempted to search for long periods using the OGLE-III photometry only. However, the time span of OGLE-III turned out to be insufficient to reliable detect periods longer 1600~days.

In order to obtain a longer time base, I extracted the OGLE-IV photometric data \citep{udalski2015, iwanek2021a} and combined it with the OGLE-III photometry. Then, I rerun the pulsation period subtraction and period search with the same procedure but in the period ranging from 40 to 2500~days. The lower limit was deliberately set to a value significantly lower than the typical value of LSP in order to detect any additional pulsational variability, that could have appeared as an alias in the long period range. I set the signal-to-noise ratio required for a Mira to be selected as a candidate to $S/N > 20$. 

Fig.~1 presents the period-luminosity (PL) diagram for all the OGLE-III LPVs, divided into three classes: OSARGs, SRVs, and Miras plotted with their most prominent period reported in the catalog \citep{soszynskietal2007}. $W_{JK}$ is an infrared Wesenheit index designed as:
\begin{equation}
    W_{JK} = K_s - 0.686(J-K_s)
\end{equation}
which is meant to be reddening independent. The $J$ and $K_s$ magnitudes come from the 2MASS survey \citep{skrutskie2006}.

I placed the 208 selected candidates on the PL diagram (Fig.~1) using the new, long periods obtained for the combined light curves. It can be noticed that these stars separate into two groups. One is lying on sequence C, where the pulsational periods of Miras are located. In this case, the residuals of the original, pulsational period are still the strongest source of periodicity detected, meaning it is unlikely to be any LSP present, therefore, I removed them from the list. The second group lies in the prolongation of sequence D, which is where the putative LSP Miras would be expected to appear. Additionally, I removed 11 objects from the sample, because the long periods identified for them turned out to be a multiplication of the pulsational period.

\section{Catalog}

The selection procedure described above resulted in a list of 108 candidates for LSP Miras. Table~1 presents a shortened version of the list with the ten first records. The full list, in a machine-readable format, is available in electronic form at the CDS via anonymous ftp to \url{cdsarc.u-strasbg.fr} (130.79.128.5) or via \url{http://cdsweb.u-strasbg.fr/cgi-bin/qcat?J/A+A/}. For each of the stars the OGLE id, the coordinates, and the periods of both fundamental mode pulsation and putative LSP are given. 

Fig.~2 presents some example light curves of the C-rich Mira LSP candidates. The O-rich candidates are presented in Fig.~3. The original light curves: both unfolded and phases-folded with the fundamental mode derived by Soszynski et al. (2007), as well as the light curve after fundamental period subtraction, folded with the putative LSP are presented.    

 The light curves of LSP candidates show a large variety of shapes from clearly periodic (e.g. OGLE-LMC-LPV-33290 and 12633) though those where the long periodicity is visible but much less prominent (e.g. OGLE-LMC-LPV-00098, 04314, and 19756) to those where the putative LSP signal is hardly visible (e.g. OGLE-LMC-LPV-47038, 52708, 87402). Most of the period-subtracted light curves still show residuals of the original, pulsational period. Some are also affected by long-term changes of magnitude (e.g. OGLE-LMC-LPV-04314) or pulsational period and amplitude (e.g. OGLE-LMC-LPV-47038). 

These factors make visual verification of the sample challenging and prone to subjective selection bias. Therefore, I decided not to include the visual inspection in the candidate selection process and not to remove any objects, even if their period-subtracted light curves do not resemble the typical LSP light curves. 

\begin{table*}
\caption{The list of the Mira LSP candidates in the LMC (first ten lines, the full list available in electronic form at the CDS.}
    \centering

    \begin{tabular}{lc c  c c}
    \hline
    ID & RA & DEC & $P_{\rm puls}$ & $P_{\rm LSP}$ \\
    \hline \\
OGLE-LMC-LPV-00098 & 04:31:03.28 & -69:34:15.3 & 323.1 & 1770.4 \\ 
OGLE-LMC-LPV-00846 & 04:37:52.76 & -69:01:47.7 & 405.5 & 1253.0 \\ 
OGLE-LMC-LPV-01018 & 04:38:51.56 & -68:24:10.4 & 517.4 & 2352.4 \\ 
OGLE-LMC-LPV-01122 & 04:39:28.91 & -67:55:35.4 & 283.0 & 1095.4 \\ 
OGLE-LMC-LPV-01427 & 04:41:03.54 & -66:42:34.0 & 435.7 & 2065.4 \\ 
OGLE-LMC-LPV-03180 & 04:47:46.50 & -69:32:13.2 & 544.5 & 1678.8 \\ 
OGLE-LMC-LPV-03327 & 04:48:11.89 & -68:16:33.8 & 258.7 & 1707.5 \\ 
OGLE-LMC-LPV-04314 & 04:50:40.53 & -68:58:19.0 & 440.8 & 1481.3 \\ 
OGLE-LMC-LPV-05548 & 04:53:10.76 & -68:57:08.3 & 897.7 & 1287.8 \\ 
OGLE-LMC-LPV-07396 & 04:55:57.68 & -69:54:48.6 & 318.5 & 1527.5 \\ 
... \\
 \\
    \hline
    \end{tabular}
    \end{table*}

\section{Discussion}

Out of the 1658 Miras in the LMC reported by \citet{soszynskietal2007}, 108 appear to show additional, periodic, long-term variability. This variability seems to be broadly consistent with the LSP that is observed for other types of LPVs (OSARGs and SRVs). This makes 7\% of the whole LMC Miras sample, which is far less than 25\% given as a lower estimate of the LSP fraction for the whole population of LPVs given by \citet{soszynskietal2007}, but still a significant fraction. It should also be noted that this number can be underestimated since some of the LSP Miras might not have passed the $S/N$ criteria.

Since no other sample of know LSP Miras is available for comparison, I use the sample of OGLE LSP SRVs in the LMC as the closest analog. Fig.~4 presents the PL relation formed by both LSP SRVs and Mira candidates in the $P$ - $W_{JK}$ plane. The relations derived by \citet{soszynski2007} for the sequence D are marked with solid and dashed lines for C- and O-rich stars respectively. The LSP Mira candidates appear to lie on the prolongation of the PL relation formed by LSP SRVs. However, the scatter in the magnitude is to large derive a meaningful slope of a PL relation. 

A recent study by \citet{iwanek2021b} shows that PL relations for Miras can be derived very accurately in the WISE \citep{wright2010} infrared filters, especially in the WISE [3.6] band. Therefore, I decided to check if this approach would also work for the putative LSP PL relation. Fig.~5 presents the LSP Mira candidates in the PL plane using WISE [3.6] magnitude. Miras seem to form an extension of the SRVs relation, but the issue of a relatively large scatter remains. 

I also compare the LSP amplitudes distribution of the SRVs and Miras. The amplitude vs. period distribution is presented in Fig.~6. The amplitudes of the LSP in Miras have been measured after prewhitening the main, pulsational variability. The Mira LSP amplitudes fall into the same range that had been found for SRVs.

Next, I investigate how the potential Mira LSP candidates differ from the rest of the Mira population in terms of their basic properties. For that purpose, I compare the distribution of the pulsational periods and amplitudes. The cumulative distribution in the period is shown in Fig.~7 and in amplitude in Fig.~8. The pulsational periods of the LSP candidates are statistically longer than those of other Miras. It seems that the putative LSP is not present in the short-period Mirasat at all. This is similar to the situation observed in OSARGs and SRVs. At the same time, the comparison of amplitudes distribution does not reveal any particular trend. I check both distributions with the Kolmogorov-Smirnov test, the difference in the distribution in periods is statistically significant, while the one in amplitudes is not.

\begin{figure*}[ht]
\centering
	\includegraphics[width=0.90\textwidth]{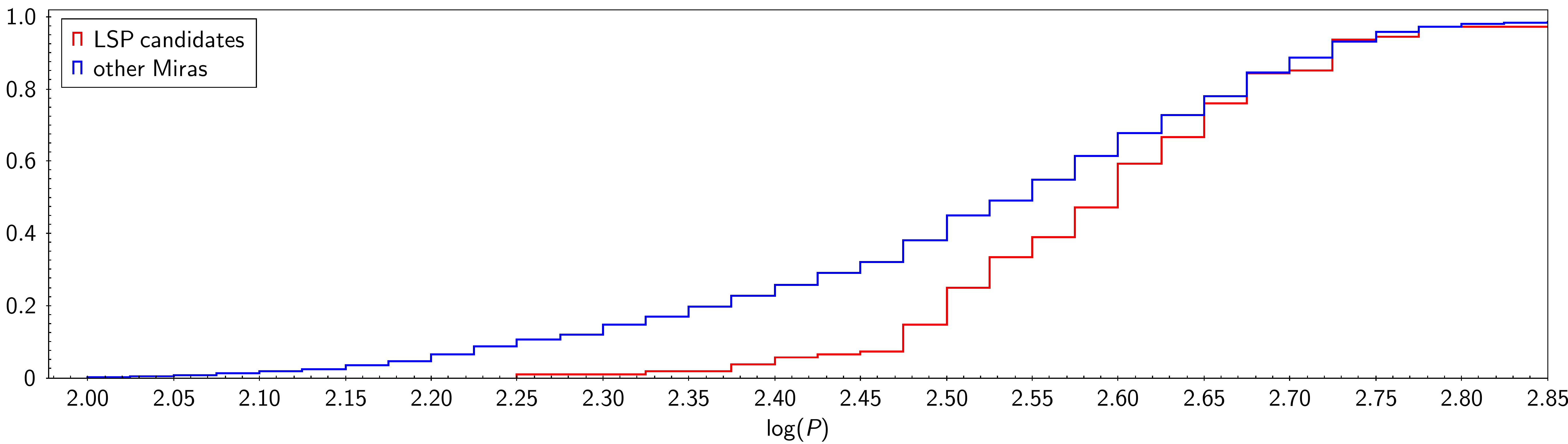} 
	\caption{The cumulative distribution in pulsational periods for the LSP candidates and other Miras.}  
	    \label{fig:mira_phist}
\end{figure*}

\begin{figure*}[ht]
\centering
	\includegraphics[width=0.90\textwidth]{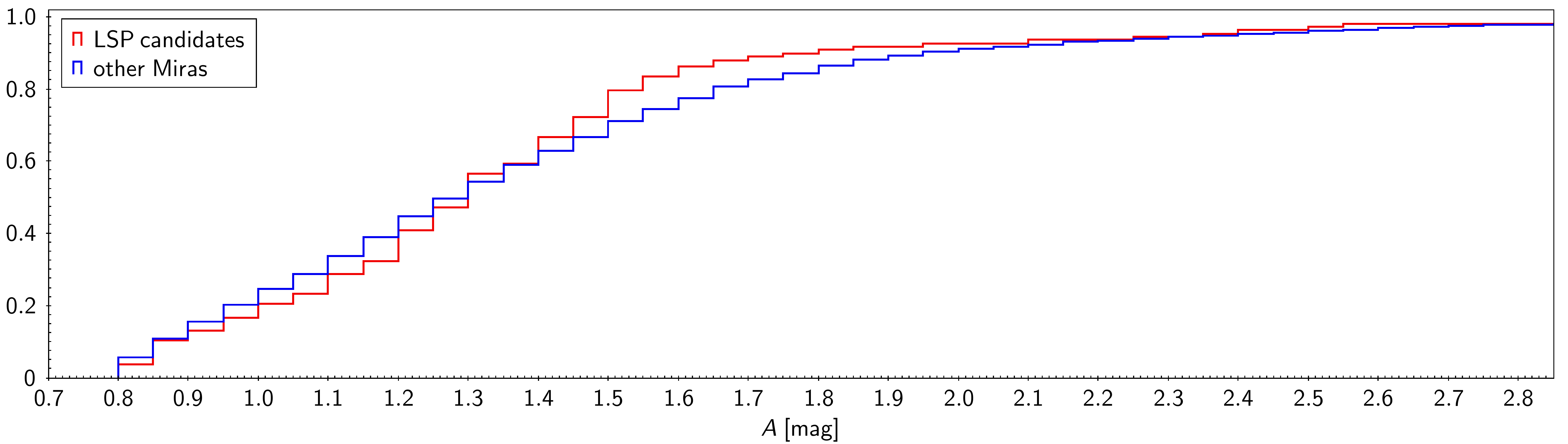} 
	\caption{The cumulative distribution in amplitudes of the pulsational variability for the LSP candidates and other Miras.}  
	    \label{fig:mira_ahist}
\end{figure*}

Another interesting aspect of the Mira LSP candidates is their C/O status. As noted in Pawlak (2021), the emergence of LSP in LPVs can be related to the transition from O-rich to C-rich, which happens over the course of the evolution on AGB. As Miras are typically evolved AGB stars, the fraction of C-rich stars among them  is generally higher than for other LPVs. As reported in \citet{soszynskietal2007}, in the entire population of OGLE LMC Miras, 1194 out of 1664 (72\%) are C-rich stars. Interestingly, this fraction is even higher for the 108 Mira LSP candidates, among which 100 (92\%) are C-rich.

\section{Summary and Conclusions}

In the OGLE catalog of the LPVs in the LMC, containing 1658 Miras, I identify 108 potential candidates for LSP Miras (7\% of the whole sample). The selection is based on the secondary period search, performed after prewhitening the primary period, related to the fundamental mode pulsation. The list of the LSP Miras candidates is provided as electronic supplementary material. The results of this work suggest that Miras may show LSPs in a similar way as SRVs and OSARGs, and that the previous lack of known LSP Miras might be an observational bias.

The putative LSP variability of the selected Miras is broadly consistent with the LSP phenomenon observed in other types of LPVs in terms of their location in the PL diagram, the amplitude of the variability, as well as the light curve shape. However, this interpretation carries a certain level of uncertainty, due to the number of factors. First, prewhitening of the fundamental mode does not remove the original, pulsational variability completely, making it harder to visually verify the putative LSP. 
The fact that Miras are high-amplitude pulsators also contributes significantly to the uncertainty of the 2MASS and Wise infrared magnitudes, which are either single-epoch or derived from a relatively small number of measurements. Finally, the putative long periods of Miras are typically longer than 1000~days, which often means that only two or three full cycles can be observed in the currently available data set.
For that reason, I chose to report the objects identified in this work as LSP candidates, rather than LSP Miras. 

I also note that Miras are known to show variations in period and amplitude. Therefore, an alternative explanation, attributing the long-term variability to these irregularities is possible. Further study, using longer time base data will be necessary to discriminate between these scenarios. Verifying if the long periods are stable over a larger number of cycles will be essential in this case.

It should also be noted, that the list  of potential candidates might be incomplete, due to the fact, that the lower-amplitude LSP candidates might not have passed the $S/N$ criterion. The sample of the Mira LSP candidates is dominated by C-rich stars to an even higher degree than the population of Miras themselves. The fact that the few O-rich stars present, seem to be mostly outliers, suggests that the putative Mira LSP population may be exclusively C-rich. The LSP candidates have statistically longer periods of the fundamental pulsation mode than the rest of the Mira sample. A similar feature is observed for OSARG and SRV LSPs, which may suggest that the mechanism behind these two phenomena is similar.

\begin{acknowledgements}
I thank the Referee for their insightful comments. I also thank Igor Soszy{\'n}ski and Patryk Iwanek for the discussion of the results.  

This study has been supported by the SONATINA grant 2020/36/C/ST9/00103 from the Polish National Science Center. 

The OGLE-III data used in this paper is publicly available via the OGLE data archive \url{http://ogledb.astrouw.edu.pl/~ogle/OCVS/}. The OGLE-IV data has been kindly provided by the OGLE Team. This work made extensive use of TOPCAT \citep{taylor2005} and VARTOOLS \citep{hartman2016}. 
\end{acknowledgements}



\begin{thebibliography}{}

\bibitem[Eddington \& Plakidis(1929)]{eddington1929} Eddington A.~S., Plakidis S., 1929, MNRAS, 90, 65. doi:10.1093/mnras/90.1.65

\bibitem[Hartman \& Bakos(2016)]{hartman2016} Hartman, J.~D. \& Bakos, G. {\'A}.\ 2016, Astronomy and Computing, 17, 1

\bibitem[Hinkle et al.(2002)]{hinkle2002} Hinkle, K.~H., Lebzelter, T., Joyce, R.~R., \& Fekel, F.~C.\ 2002, \aj, 123, 1002 

\bibitem[Houk(1963)]{houk1963} Houk, N.\ 1963, \aj, 68, 253

\bibitem[Iwanek et al.(2021a)]{iwanek2021a} Iwanek P., Koz{\l}owski S., Gromadzki M., Soszy{\'n}ski I., Wrona M., Skowron J., Ratajczak M., et al., 2021, ApJS, 257, 23

\bibitem[Iwanek et al.(2021b)]{iwanek2021b} Iwanek P., Soszy{\'n}ski I., Koz{\l}owski S., 2021, ApJ, 919, 99

\bibitem[\protect\citeauthoryear{McDonald \& Trabucchi}{2019}]{mcdonald2019} McDonald I., Trabucchi M., 2019, MNRAS, 484, 4678

\bibitem[O'Connell(1933)]{oconnell1933} O'Connell, D.~J.~K.\ 1933, Harvard College Observatory Bulletin, 893, 19 

\bibitem[Pawlak(2021)]{pawlak2021} Pawlak M., 2021, A\&A, 649, A110

\bibitem[Payne-Gaposchkin(1954)]{paynegaposchkin1954} Payne-Gaposchkin, C.\ 1954, Annals of Harvard College Observatory, 113, 189 

\bibitem[Percy \& Au(1999)]{percy1999a} Percy J.~R., Au W.~W.-Y., 1999, PASP, 111, 98

\bibitem[Percy \& Colivas(1999)]{percy1999b} Percy J.~R., Colivas T., 1999, PASP, 111, 94

\bibitem[Plakidis(1932)]{plakidis1932} Plakidis, S.,\ 1932, MNRAS, 92, 460

\bibitem[Saio et al.(2015)]{saio2015} Saio, H., Wood, P.~R., Takayama, M., \& Ita, Y.\ 2015, \mnras, 452, 3863 

\bibitem[Skrutskie et al.(2006)]{skrutskie2006}Skrutskie et al. 2006, AJ, 131, 1163

\bibitem[Soszy{\'n}ski et al.(2005)]{soszynski2005} Soszy{\'n}ski I., Udalski A., Kubiak M., Szyma{\'n}ski M.~K., Pietrzy{\'n}ski G., {\.Z}ebru{\'n} K., Szewczyk O., et al., 2005, AcA, 55, 331

\bibitem[Soszy{\'n}ski et al.(2007)]{soszynskietal2007} Soszynski, I., Dziembowski, W.~A., Udalski, A., et al.\ 2007, \actaa, 57, 201 

\bibitem[Soszy{\'n}ski(2007)]{soszynski2007} Soszy{\'n}ski, I.\ 2007, \apj, 660, 1486 

\bibitem[Soszy{\'n}ski and Udalski(2014)]{soszynski2014} Soszy{\'n}ski, I. \& Udalski, A. 2014, \apj

\bibitem[Soszy{\'n}ski et al.(2021)]{soszynski2021} Soszy{\'n}ski I., Olechowska A., Ratajczak M., Iwanek P., Skowron D.~M., Mr{\'o}z P., Pietrukowicz P., et al., 2021, ApJL, 911, L22

\bibitem[Sterne \& Campbell(1937)]{sterne1937} Sterne T.~E., Campbell L., 1937, AnHar, 105, 459

\bibitem[Taylor(2005)]{taylor2005} Taylor M.~B., 2005, ASPC, 347, 29

\bibitem[Trabucchi et al.(2017)]{trabucchi2017} Trabucchi M., Wood P.~R., Montalb{\'a}n J., Marigo P., Pastorelli G., Girardi L., 2017, ApJ, 847, 139

\bibitem[Udalski(2003)]{udalski2003} Udalski, A.\ 2003, \actaa, 53, 291 

\bibitem[Udalski et al.(2015)]{udalski2015} Udalski A., Szyma{\'n}ski M.~K., Szyma{\'n}ski G., 2015, AcA, 65, 1

\bibitem[Wood et al.(1999)]{wood1999} Wood, P.~R., Alcock, C., Allsman, R.~A., et al.\ 1999, Asymptotic Giant Branch Stars, 191, 151 

\bibitem[Wood(2000a)]{wood2000a} Wood, P.~R.\ 2000, IAU Colloq.~176: The Impact of Large-Scale Surveys on Pulsating Star Research, 203, 379 

\bibitem[Wood(2000b)]{wood2000b} Wood, P.~R.\ 2000, \pasa, 17, 18 
\bibitem[Wood et al.(2004)]{wood2004} Wood, P.~R., Olivier, E.~A., \& Kawaler, S.~D.\ 2004, \apj, 604, 800

\bibitem[Wright et al.(2010)]{wright2010} Wright E.~L., Eisenhardt P.~R.~M., Mainzer A.~K., Ressler M.~E., Cutri R.~M., Jarrett T., Kirkpatrick J.~D., et al., 2010, AJ, 140, 1868

\bibitem[Zechmeister \& K{\"u}rster(2009)]{zechmeister2009} Zechmeister, M. \& K{\"u}rster, M.\ 2009, \aap, 496, 577

\end{thebibliography}
\end{document}